\PassOptionsToPackage{hidelinks}{hyperref}
\documentclass[12pt]{article}

\usepackage[utf8]{inputenc}
\usepackage[english]{babel}
\usepackage[babel]{csquotes}
\usepackage[
    backend=biber,
    sorting=none,
    style=numeric,
    citestyle=numeric-comp,
    maxcitenames=2,
    defernumbers=true
]{biblatex}
\usepackage{graphicx}
\usepackage{amsmath}
\usepackage{geometry}
\usepackage{orcidlink}
\usepackage{titlesec}
\usepackage{url}
\usepackage{hyperref}
\usepackage{bookmark}
\usepackage{float} 
\usepackage{authblk}
\usepackage{setspace}
\usepackage{xcolor}
\usepackage{comment}

\begin{document}

\titleformat{\section}{\large\bfseries}{\thesection.}{1em}{}

\vspace*{-1.5cm}
\noindent\textit{Article} \\[1em]

\begin{center}
    {\LARGE \textbf{Watershed vs. Region Growing for Individual Tree Segmentation from Airborne LiDAR: An Urban Case Study in Bologna} \\[1em]}
\end{center}

\noindent
\textbf{Aldo Canfora}\,\orcidlink{0009-0006-2770-4413},
\textbf{Tommaso Rondini}\,\orcidlink{0009-0001-2867-212X},
\textbf{Matteo Falcioni},
\textbf{Mirko Degli Esposti}\,\orcidlink{0000-0003-0316-3449}

\begin{center}
  Department of Physics and Astronomy, University of Bologna, Via Irnerio 46, 40126 Bologna, Italy
\end{center}
\begin{center}
  July 2026
\end{center}
\begin{center}
   \url{https://github.com/physycom/Vegetation_Segmentation}
\end{center}

\begin{abstract}
\noindent
We present a LiDAR-based pipeline for the segmentation and structural characterisation of tall urban vegetation in Bologna.
Starting from airborne LiDAR point clouds previously classified as high vegetation by a Random Forest model, we implement and compare two individual-tree segmentation strategies: a watershed algorithm applied to the Canopy Height Model and a point-wise region growing algorithm operating directly on the three-dimensional cloud.
Over an area of 12 tiles covering the \emph{Talea} district, the watershed method detects a larger number of trees ($7589$), dominated by short trees with a height peak around 5 meters, whereas region growing yields fewer trees ($6432$) but retains the highest returns, yielding individuals above 40 meters, reflecting the absence of a smoothing step in the latter.
We further confront the segmentation results with the municipal Open Data catalogue \emph{Alberi in manutenzione}: on a sample tile roughly half of the LiDAR-detected trees turn out to be absent from the catalogue, several catalogued positions correspond to locations where no tree is observed, and the majority of the height records date back about two decades, which makes the dataset unsuitable as a ground truth reference.
From the segmented trees we extract height and crown radius and use them to compute preliminary vegetation indicators, namely above-ground biomass and carbon storage through allometric relations and annual pollen production through species-specific inflorescence coefficients.
All results are collected in an interactive per-tree metadata map.
The pipeline is modular and can be recalibrated as improved classifications, field campaigns, or complementary sensing technologies become available, providing a scalable basis for data-driven urban greenery planning.
\end{abstract}

\section{Introduction}
\begin{figure}[H]
    \centering
    \includegraphics[width=0.6\textwidth]{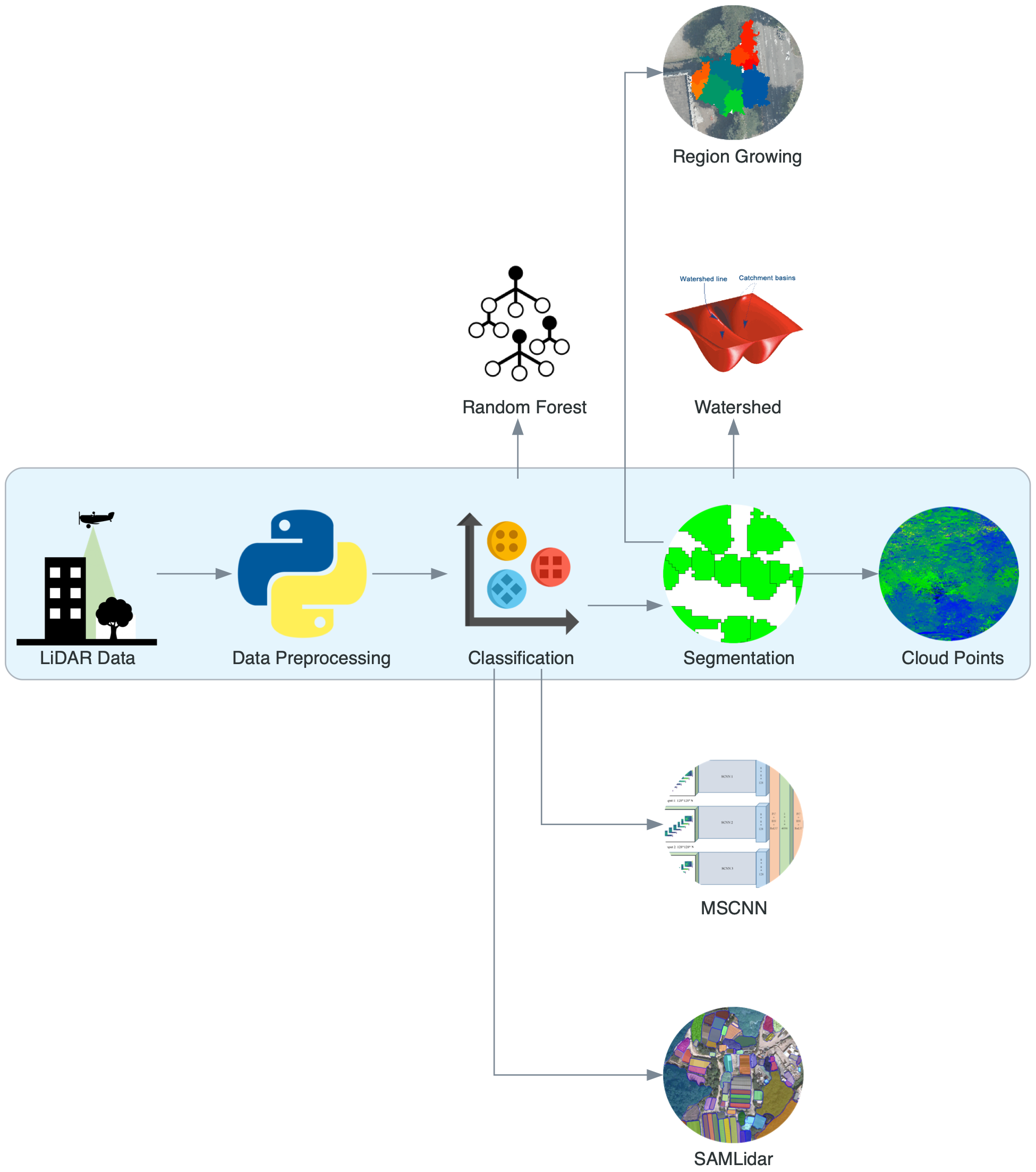}
    
    \caption{Workflow pipeline, from data to feature extraction.}
    \label{fig:pipeline}
\end{figure}

This article presents a LiDAR-based methodology for the segmentation and analysis of tall urban vegetation in Bologna.
Starting from classified airborne LiDAR data, we apply and compare two tree segmentation techniques --watershed and region growing-- highlighting their respective strengths and limitations.
We also analyze the municipal Open Data on trees, revealing significant inconsistencies and missing records when compared to the LiDAR-derived data.
Using the segmented individual trees, we extract structural features such as height and crown radius, which form the basis for preliminary vegetation indicators, namely carbon sequestration estimates via allometric models and annual pollen production estimates via species-specific inflorescence coefficients.
Due to the absence of reliable ground truth data, this work serves as a foundation for future developments, offering adaptable tools to support evolving municipal strategies on urban greenery.
The overall pipeline for the point cloud analysis is summarised in Fig.~\ref{fig:pipeline}.

\section{Random Forest Classifier}
The segmentation methods explored in this study assume that the input data have already been classified as high vegetation.
Classification was performed using a Random Forest model trained and validated on similar airborne LiDAR data~\cite{lohani2017airborne}\cite{song2015decision}.
Further details are available in a previous report (see also~\cite{amslaurea33368}).

It is worth noting that, if and when more accurate or updated vegetation classifications become available, they can be directly substituted into the pipeline.
The downstream segmentation and analysis steps remain unchanged, ensuring the modularity and adaptability of the proposed approach.

\section{Open Data - Alberi in Manutenzione}
The Open Data Comune di Bologna portal, maintained by the Municipality of Bologna, offers a comprehensive repository of datasets related to various aspects of the urban environment.
Among these, the dataset titled \emph{Alberi in manutenzione} constitutes a key resource for this study, providing detailed records on trees currently under municipal management.
The dataset can be accessed at the Open Data Bologna portal:
\textcolor{blue}{\href{https://opendata.comune.bologna.it/explore/dataset/alberi-manutenzioni/information/?disjunctive.cl_h&disjunctive.dimora&disjunctive.classe&disjunctive.zona_di_prossimita&disjunctive.area_statistica}{alberi in manutenzione}}. 

\begin{figure}[H]
    \centering
    \includegraphics[width=0.8\textwidth]{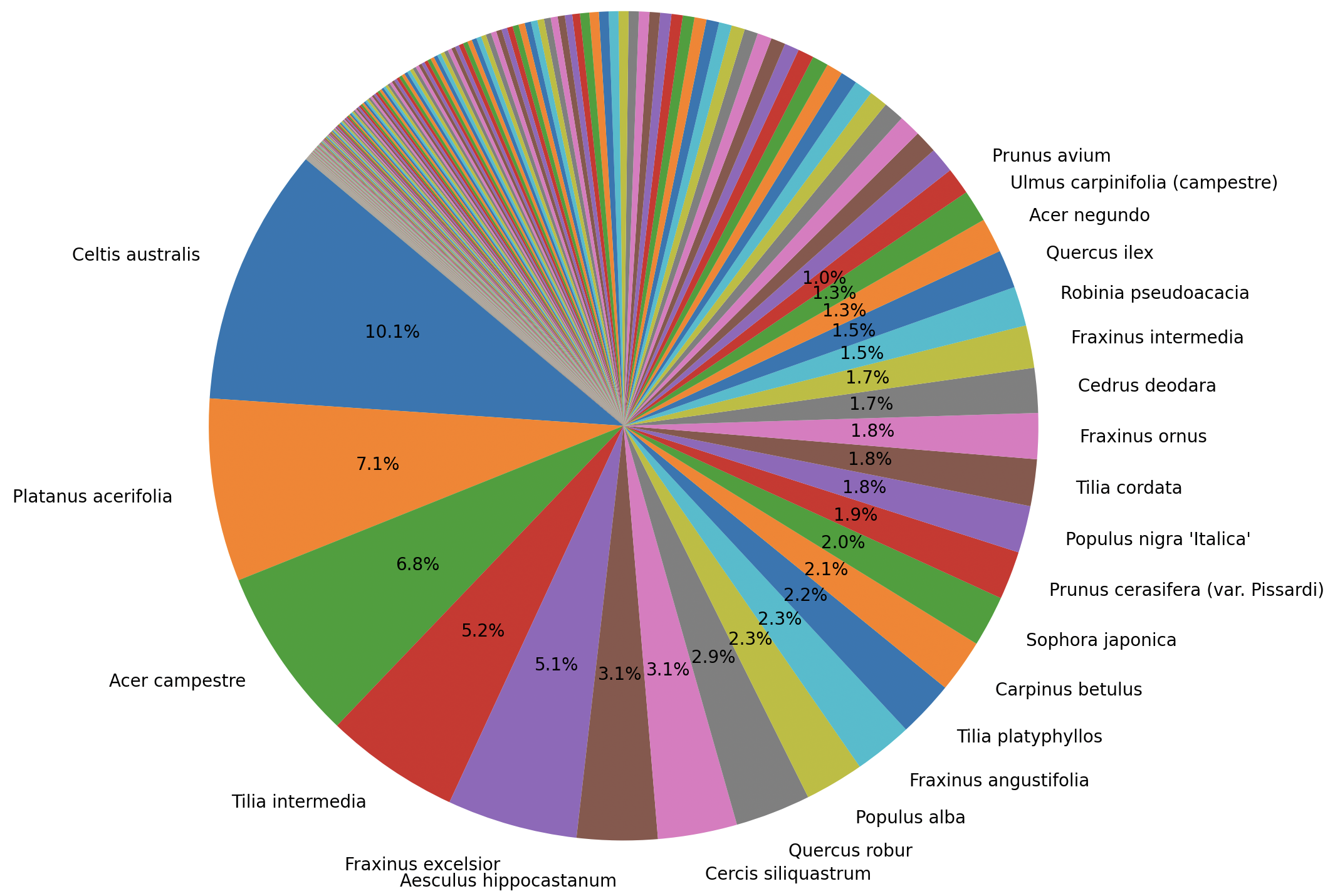}
    \caption{Percentage distribution of tree species in Bologna.}
    
    \label{fig:opendata_errors_1}
\end{figure}

\begin{figure}[H]
    \centering
    \includegraphics[width=0.8\textwidth]{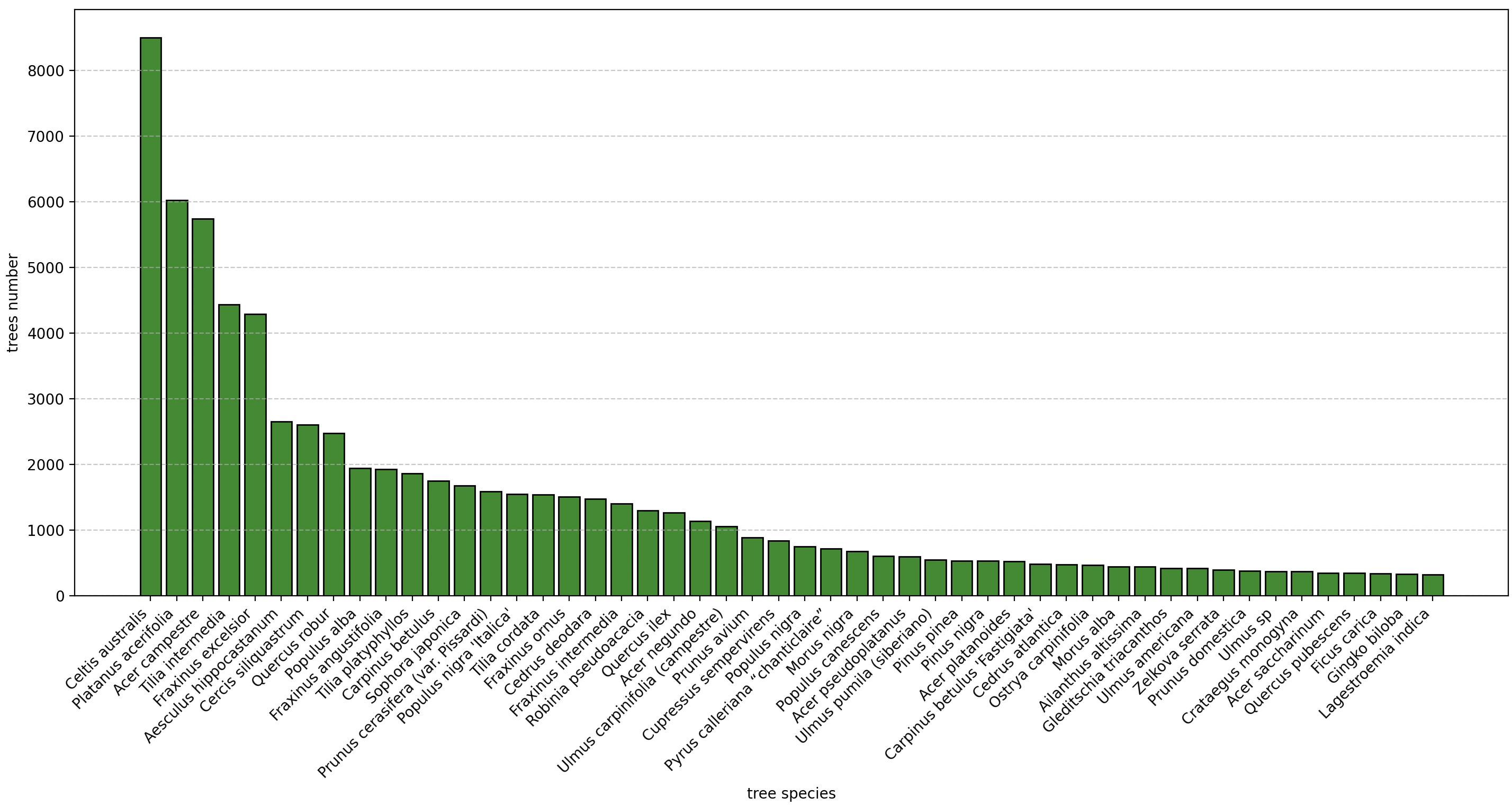}
    \caption{Absolute distribution of tree species in Bologna. }
    
    \label{fig:opendata_errors_2}
\end{figure}

\begin{figure}[H]
    \centering
    \includegraphics[width=0.8\textwidth]{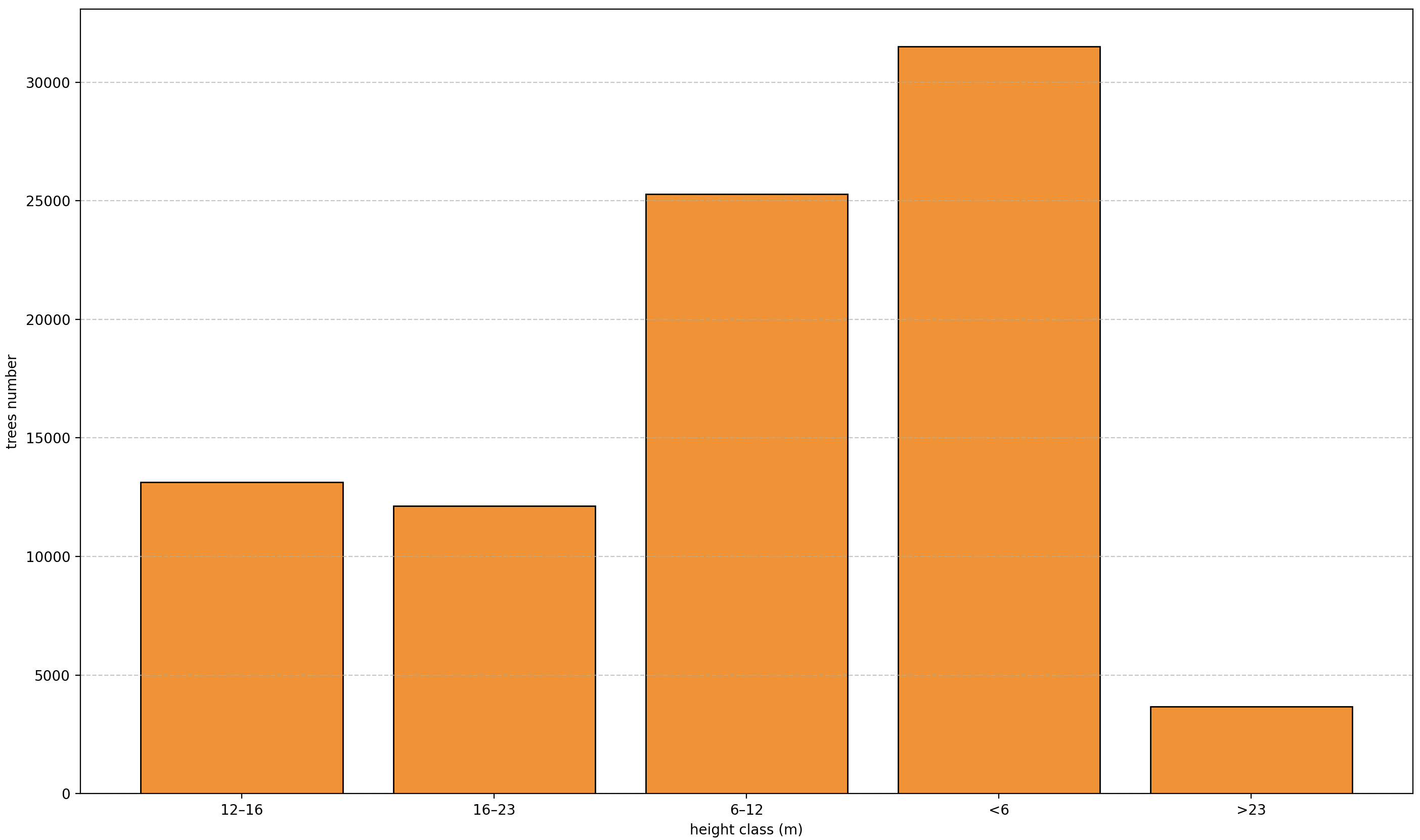}
    \caption{Distribution in tree height classes in Bologna.}
    
    \label{fig:opendata_errors_3}
\end{figure}

\begin{figure}[H]
    \centering
    \includegraphics[width=0.8\textwidth]{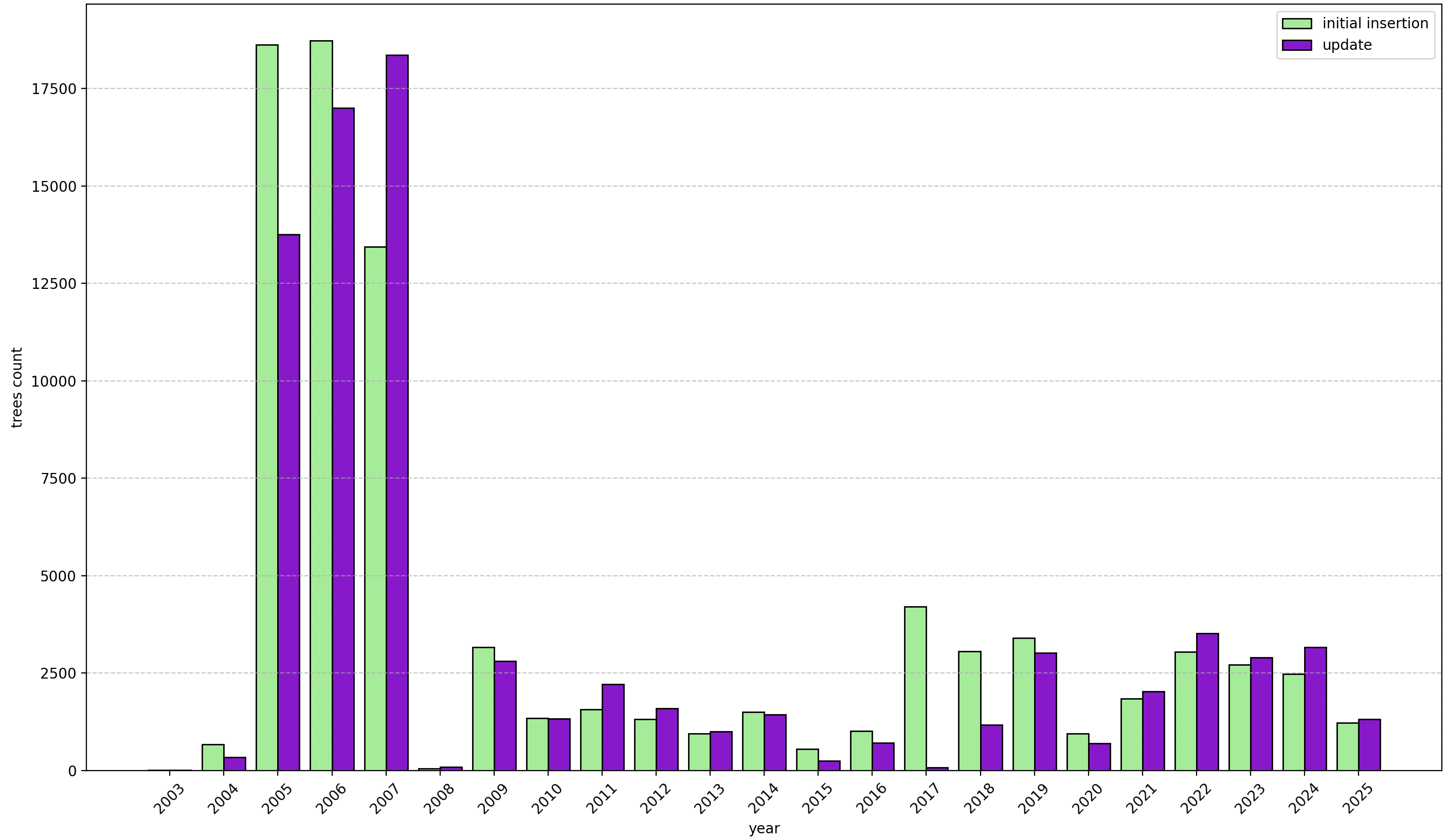}
    \caption{Distribution of initial insertion vs. update over the years of Bologna trees in the dataset.}
    
    \label{fig:opendata_errors_5}
\end{figure}

As illustrated in Fig.~\ref{fig:opendata_errors_1} and Fig.~\ref{fig:opendata_errors_2}, the urban forest of Bologna is dominated by a limited number of tree species.
Fig.~\ref{fig:opendata_errors_3} presents the distribution of trees across height classes.
However, these data may be outdated, as the majority of records originate from surveys conducted approximately two decades ago, as evidenced in Fig.~\ref{fig:opendata_errors_5}.

\begin{figure}[H]
    \centering
    \includegraphics[width=0.8\textwidth]{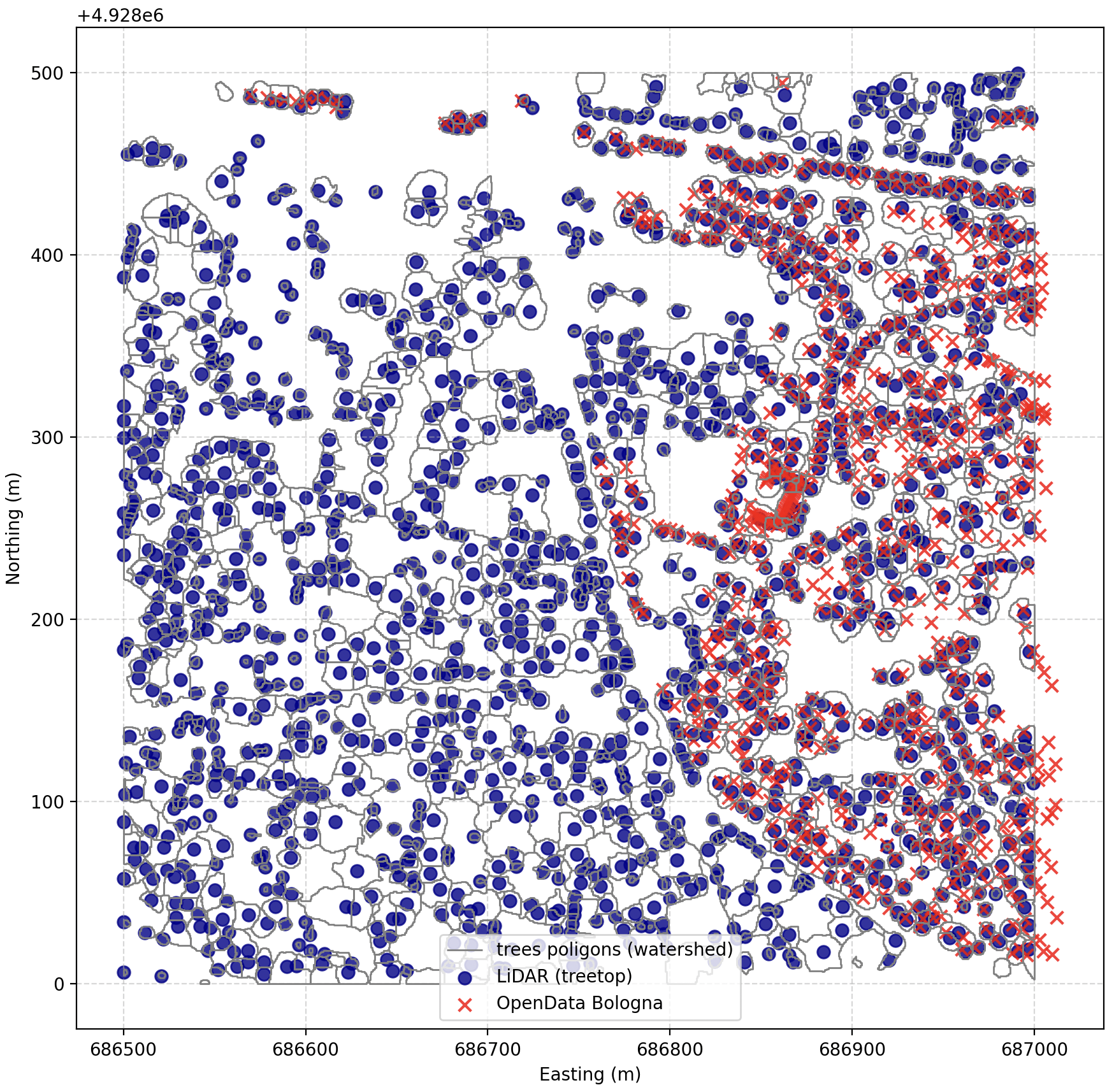}
    \caption{Difference between trees catalogued by Open Data (in red) and trees segmented with LiDAR data (in blue) with watershed method.}
    
    \label{fig:opendata_errors_6}
\end{figure}

As shown in Fig.~\ref{fig:opendata_errors_6}, even when considering only a single tile, the Open Data dataset exhibits significant gaps:
approximately half of the trees detected and segmented from the LiDAR data are not recorded in the Open Data catalog.

\begin{figure}[H]
    \centering
    \includegraphics[width=0.425\textwidth]{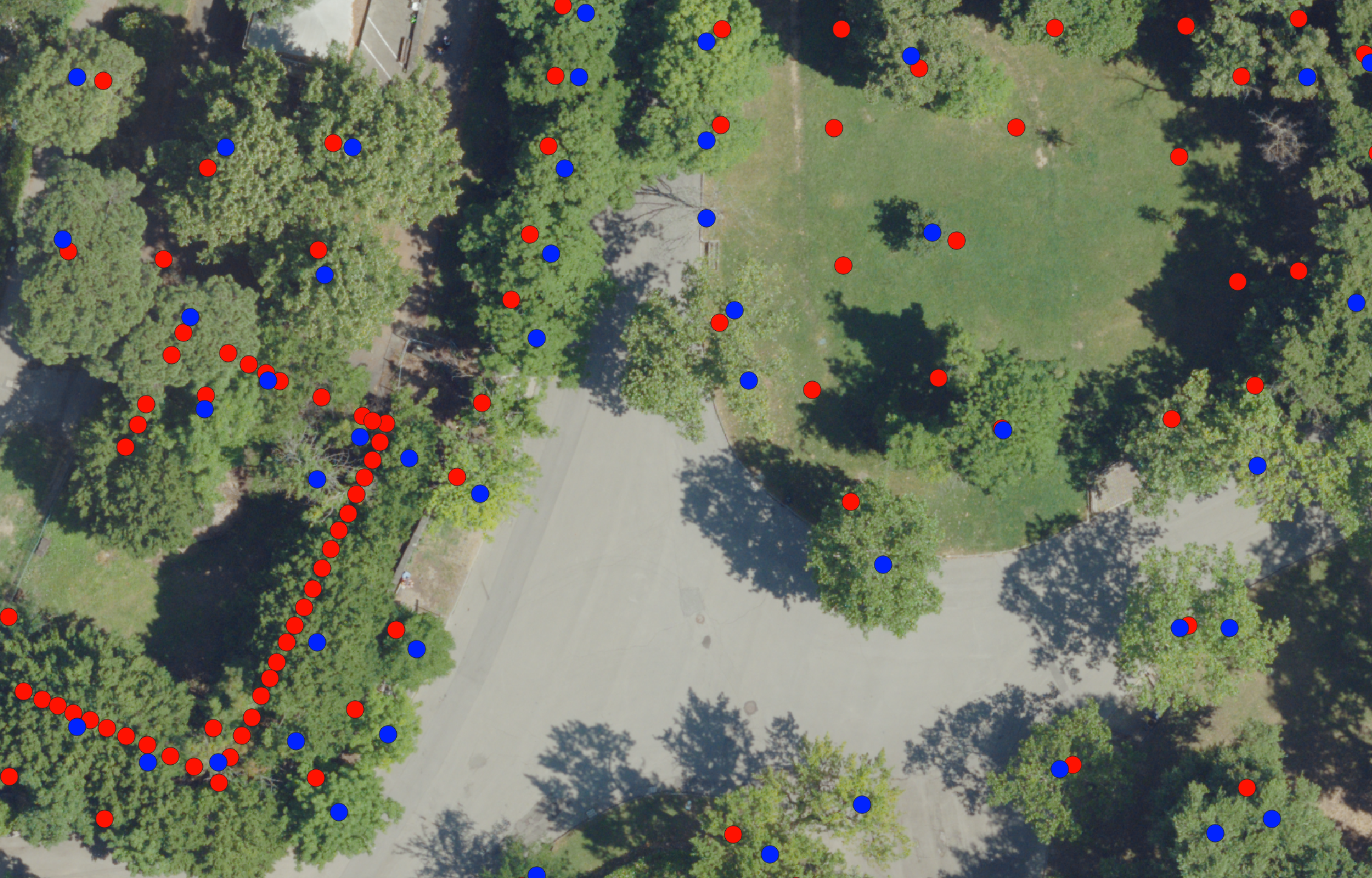}
    \includegraphics[width=0.4\textwidth]{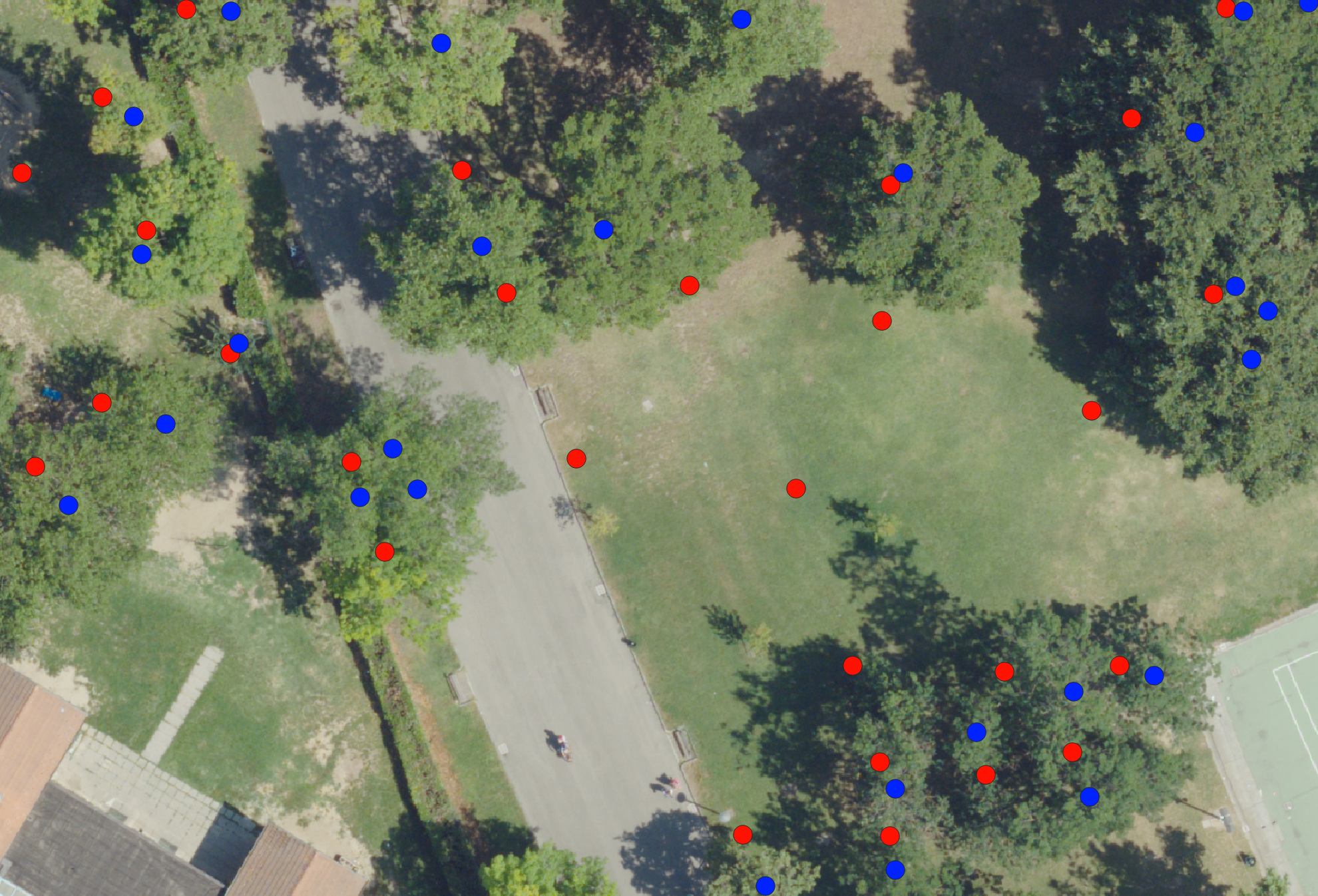}
    \caption{An example of some anomalies in Open Data in Bologna, there are some Open Data geopoints (in red) where there are no trees. Treetop from LiDAR in blue.}
    
    \label{fig:opendata_errors_7}
\end{figure}

Additionally, Fig.~\ref{fig:opendata_errors_7} highlights numerous anomalies in the Open Data records,
including instances of trees that are no longer present --possibly due to removal-- or general inaccuracies within the dataset.
These issues, however, can be effectively addressed through the integration of regularly updated aerial LiDAR acquisitions.

\begin{figure}[H]
    \centering
    \includegraphics[width=0.8\textwidth]{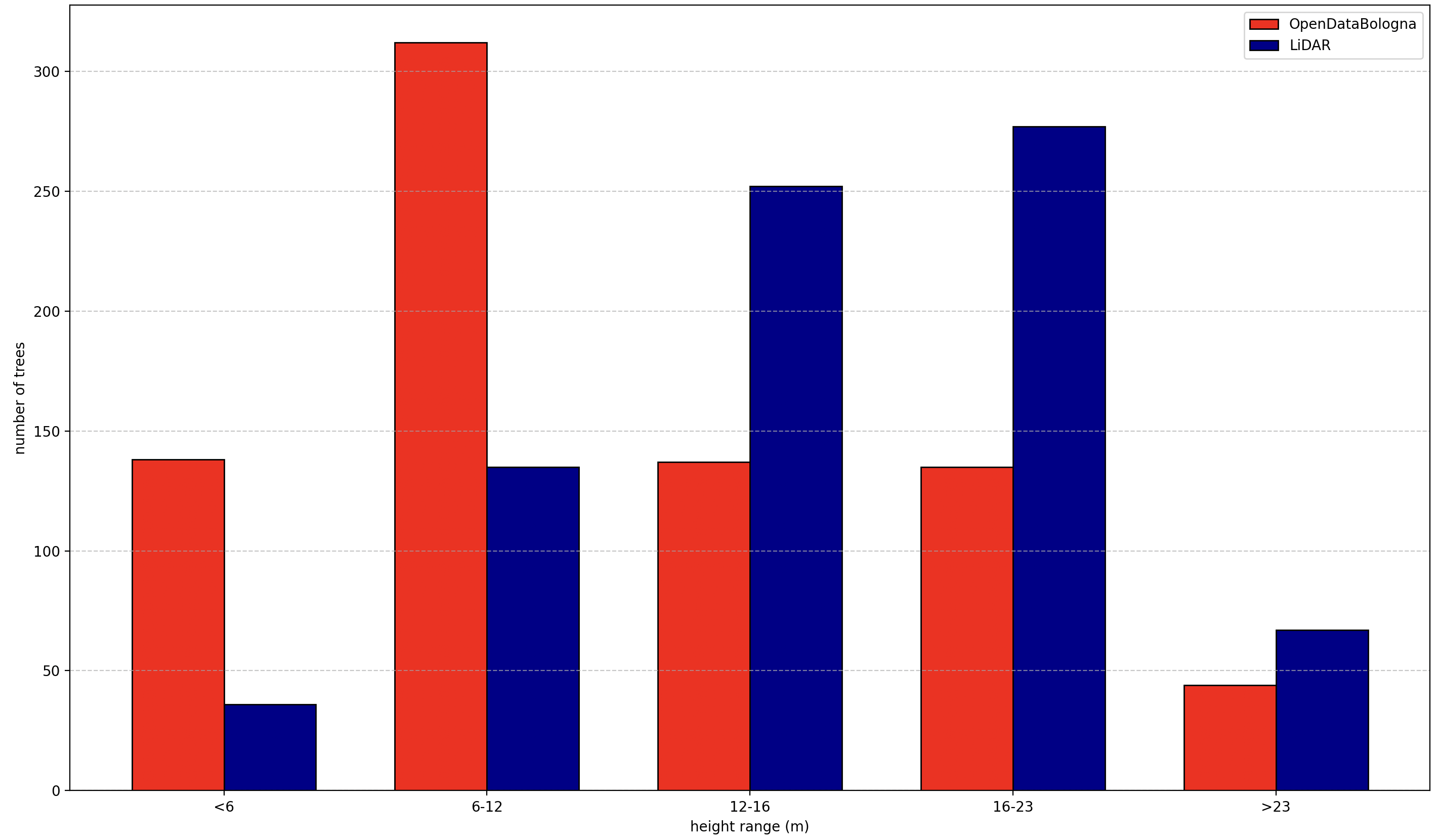}
    \caption{Height comparison between Open Data and LiDAR. The lidar heights are computed with watershed method (see section~\ref{sec:watershed}).}
    
    \label{fig:opendata_errors_4}
\end{figure}

The availability of annual aerial LiDAR data enables the extraction of key morphological features for individual trees, such as exact position, canopy radius and tree height.
As shown in Fig.~\ref{fig:opendata_errors_4}, the height information --originally provided in discrete height classes within the Open Data catalogue-- has been refined using LiDAR measurements for trees located in a selected area of Bologna.
The comparison reveals that some trees likely have been removed, as indicated by significantly lower or missing height values, while others appear to have grown.
Notably, the majority of the original height records date back around 20 years, highlighting the importance of periodic updates for maintaining data accuracy.

These observations clearly indicate that the Alberi in manutenzione dataset cannot be reliably used for training or evaluating single-tree extraction algorithms.
The presence of significant temporal gaps, missing records, and inconsistencies --such as trees no longer present, unregistered individuals, or outdated height measurements-- undermine the dataset's suitability as a ground-truth reference.
Consequently, the two extraction methods described below must be evaluated independently of this dataset.

We now turn to the single tree segmentation algorithms.

\section{Segmentation}
Tree segmentation is the process of identifying and isolating individual trees from remote sensing data, particularly from airborne LiDAR point clouds.
This technique plays a critical role in tree inventory, ecological monitoring, urban planning, and biomass estimation, as it enables the extraction of key structural metrics such as tree height, crown diameter, and spatial distribution.

In this study, we investigate and compare two widely used approaches for individual tree segmentation: watershed and region growing.
Each method presents distinct advantages in handling various canopy architectures and topographic conditions, which we evaluate within the context of our dataset.

It is important to note, however, that our analysis is based on aerial LiDAR data, which --despite its broad coverage and relatively high resolution-- has intrinsic limitations.
In particular, aerial LiDAR may struggle to detect understory vegetation or to precisely capture crown morphology in densely vegetated areas.
For a more detailed characterization of individual trees --including species identification, crown architecture, and health status-- data from terrestrial (mobile) LiDAR acquisitions or hyperspectral LiDAR sensors would be significantly more informative, as they provide complementary perspectives and spectral signatures not accessible from aerial platforms alone.

\subsection{Watershed}
\label{sec:watershed}
The watershed algorithm is a region-based segmentation method inspired by the hydrological concept of watersheds in topography.
It interprets a grayscale image as a three-dimensional surface, where pixel intensities are treated as elevation values:
high-intensity pixels represent ridges or peaks, while low-intensity pixels represent valleys or catchment basins.

\begin{figure}[H]
    \centering
    \includegraphics[width=0.45\textwidth]{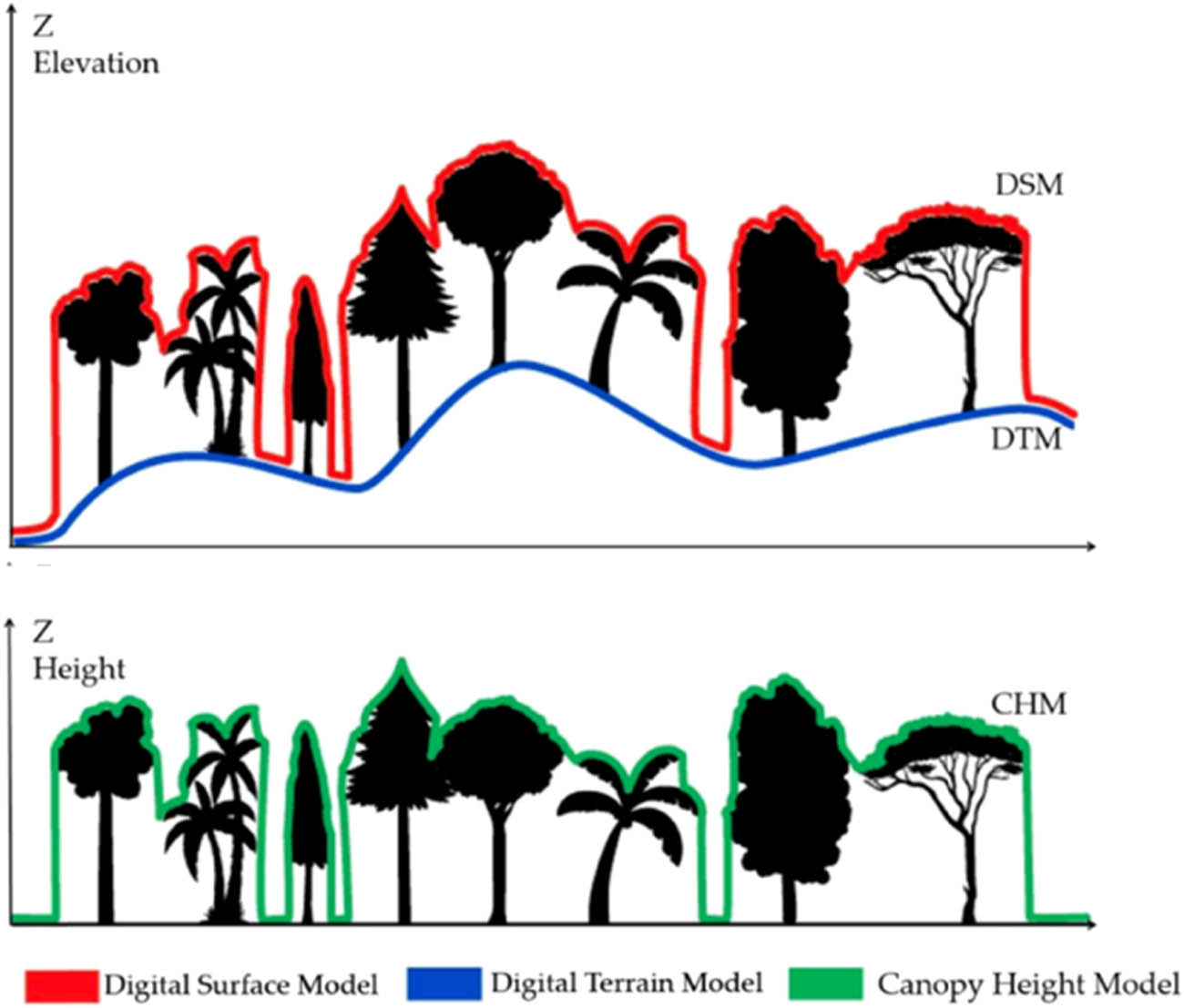}
    \includegraphics[width=0.45\textwidth]{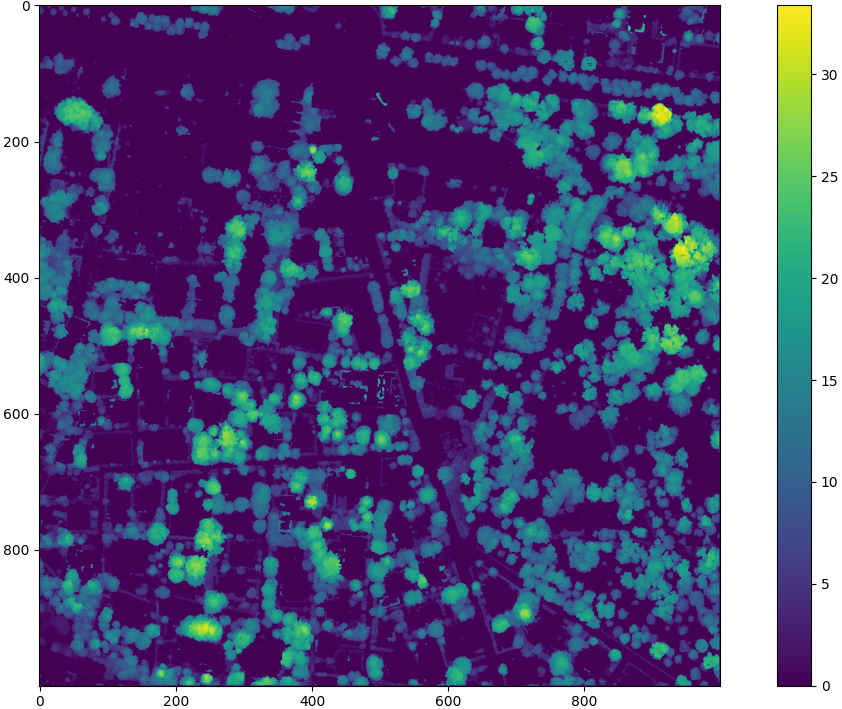}
    \caption{Left: Canopy Height Model (CHM) representation \cite{chm_image}.
    Right: CHM for a tile in Bologna, the scale represents heights in meters.}
    
    \label{fig:chm}
\end{figure}

In the context of tree segmentation, the input to the watershed algorithm is typically a \textit{Canopy Height Model} (CHM) --a raster representing the vertical height of vegetation (primarily trees) above ground level.
The CHM is computed by subtracting the Digital Terrain Model (DTM) from the Digital Surface Model (DSM) (see Fig.~\ref{fig:chm}), restricted to the vegetation class:
\[
\text{CHM} = DSM_v - DTM_v,
\]
where \( DSM_v \) and \( DTM_v \) are the DSM and DTM values associated with vegetation points only.

\begin{figure}[H]
    \centering
    \includegraphics[width=0.8\textwidth]{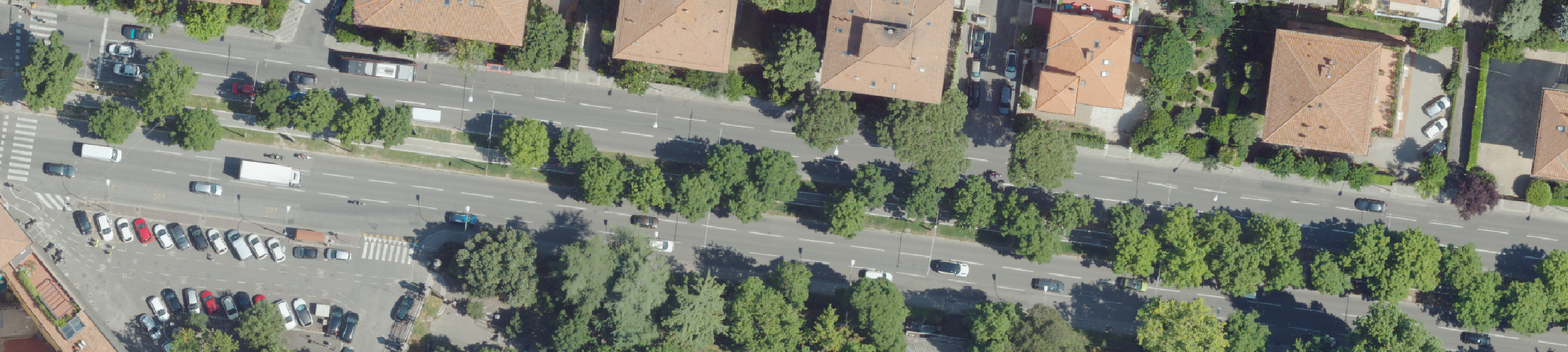}
    \includegraphics[width=0.8\textwidth]{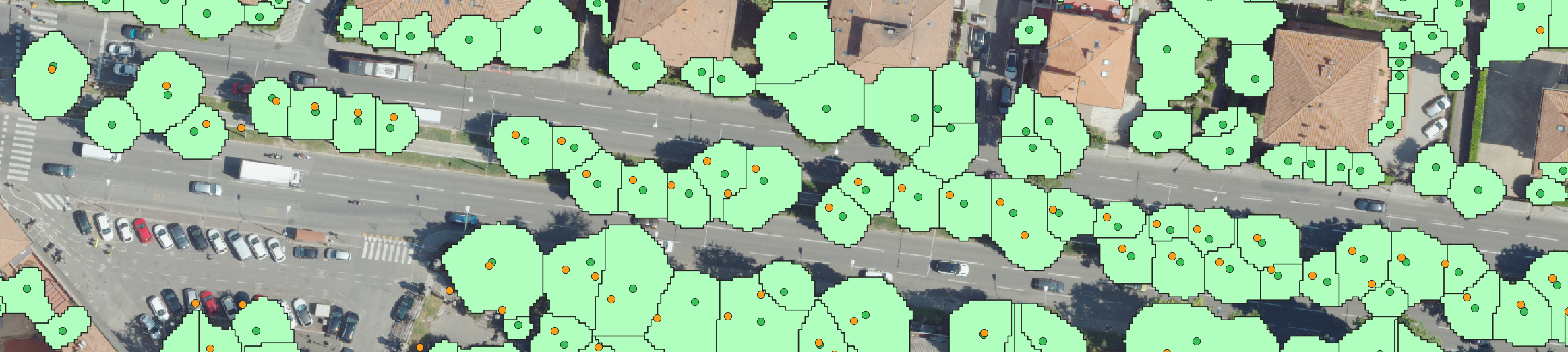}
    \caption{Example of tree segmentation using the watershed algorithm.
    The boundaries of individual tree crowns are highlighted in light green, treetops are in darkgreen and geopoints from Opendata are in orange.}
    
    \label{fig:watershed_segmentation}
\end{figure}

For individual tree crown delineation, the watershed algorithm uses the local maxima of the CHM --corresponding to tree tops-- as seed points or markers.
From each marker, the algorithm simulates the flooding of the topographic surface:
water spreads outward from the local maxima, filling basins until it encounters water from adjacent basins.
The boundaries where these basins meet define the watershed lines, which are interpreted as the borders between adjacent tree crowns.
An example is shown in Fig.~\ref{fig:watershed_segmentation}.

This method is particularly effective in settings where individual tree crowns are distinguishable in the CHM, such as in managed urban forests or sparsely vegetated areas.
However, its performance is sensitive to the quality of the CHM, the accuracy of local maxima detection, and the management of over-segmentation --especially in cases of dense canopy structures, noise, or closely spaced trees.
Appropriate pre-processing and post-processing steps are often required to mitigate these limitations.

\begin{figure}[H]
    \centering
    \includegraphics[width=0.96\textwidth]{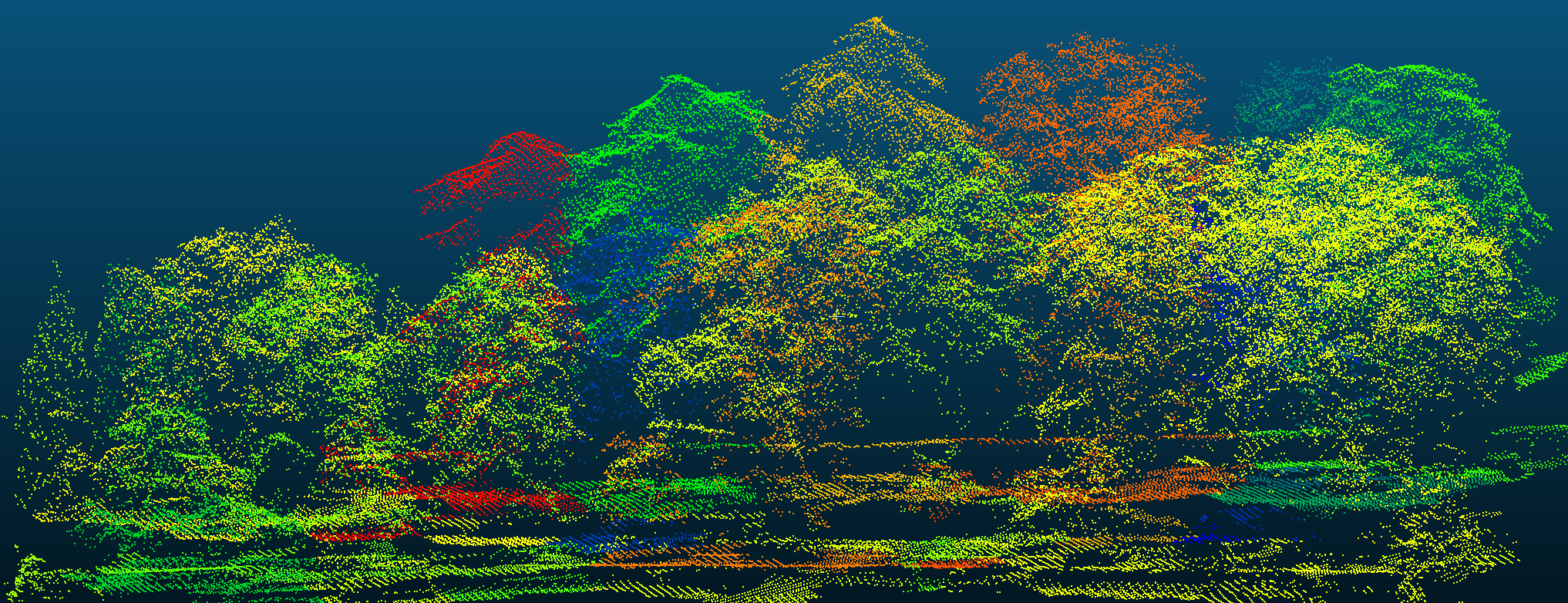}
    \caption{Tree cloud points segmented using watershed.}
    
    \label{fig:cloud_points_watershed}
\end{figure}

\begin{figure}[H]
    \centering
    \includegraphics[width=0.32\textwidth]{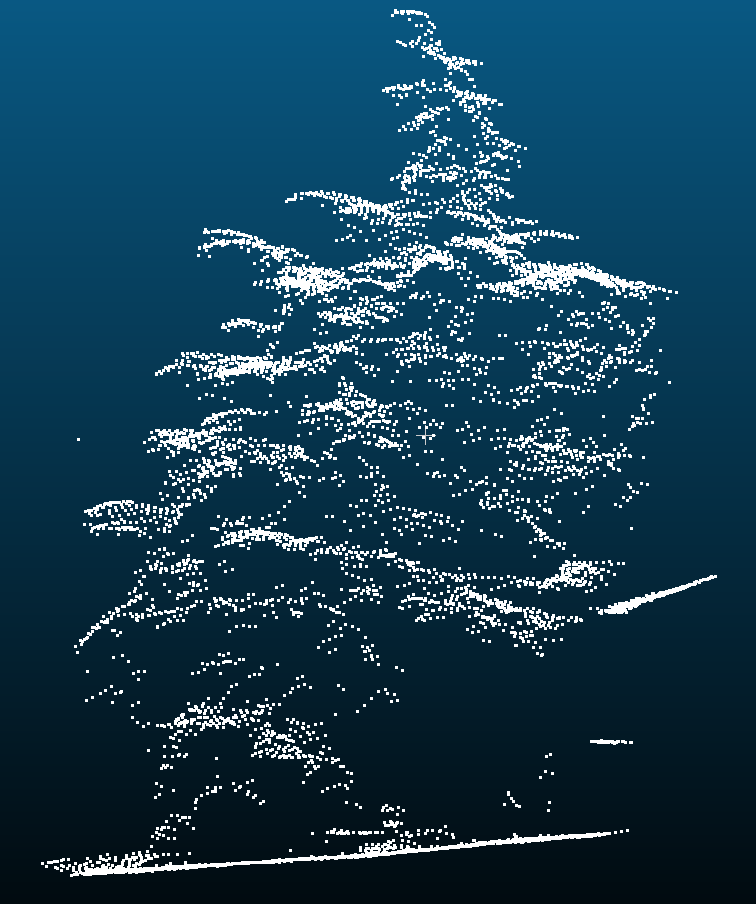}
    \includegraphics[width=0.3035\textwidth]{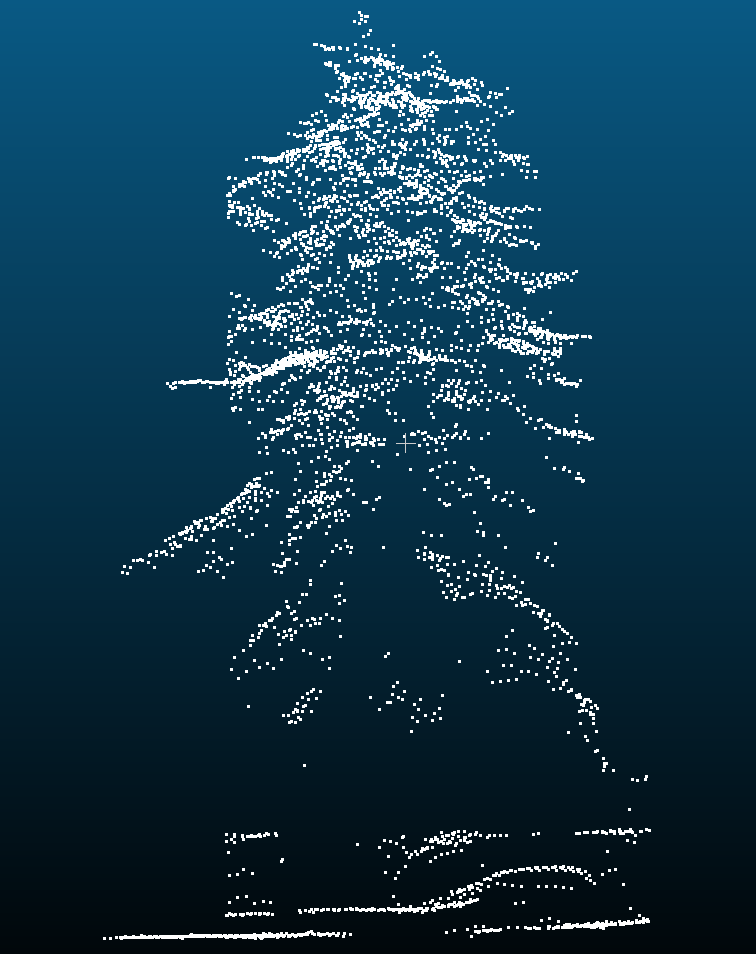}
    \includegraphics[width=0.32\textwidth]{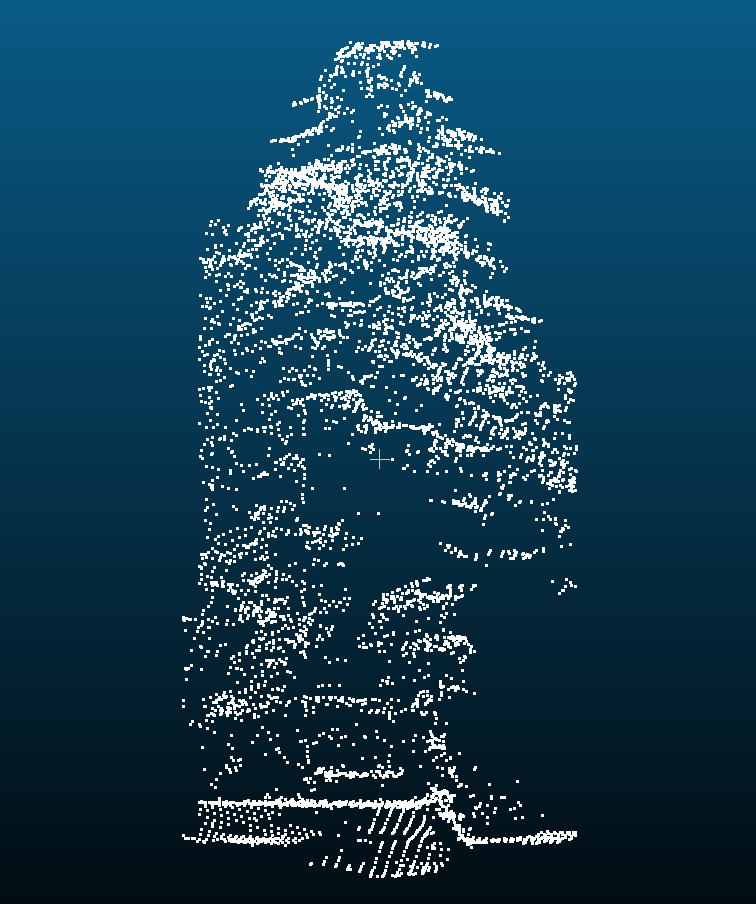}
    \caption{Single tree cloud points segmented using watershed.}
    
    \label{fig:single_cloud_points_watershed}
\end{figure}

Once the individual tree crowns are delineated using the watershed segmentation on the 
CHM, the resulting polygonal boundaries can be used to spatially filter the original 
LiDAR point cloud.
By overlaying the segmented crown boundaries onto the point cloud 
data, it is possible to extract all 3D points that fall within each individual tree's 
extent (Fig.~\ref{fig:cloud_points_watershed}).
This enables the isolation of point cloud subsets corresponding to single trees, 
facilitating detailed structural analysis at the tree level, such as heights and crown 
radius (Fig.~\ref{fig:single_cloud_points_watershed}).

\subsection{Region Growing}
The alternative approach we implemented is based on the method proposed by Li et al.~\cite{reg_grow}, which segments individual trees by assigning each LiDAR point to a tree crown based on the spatial relationships with its neighbors.

\begin{figure}[H]
    \centering
    \includegraphics[width=0.4\textwidth]{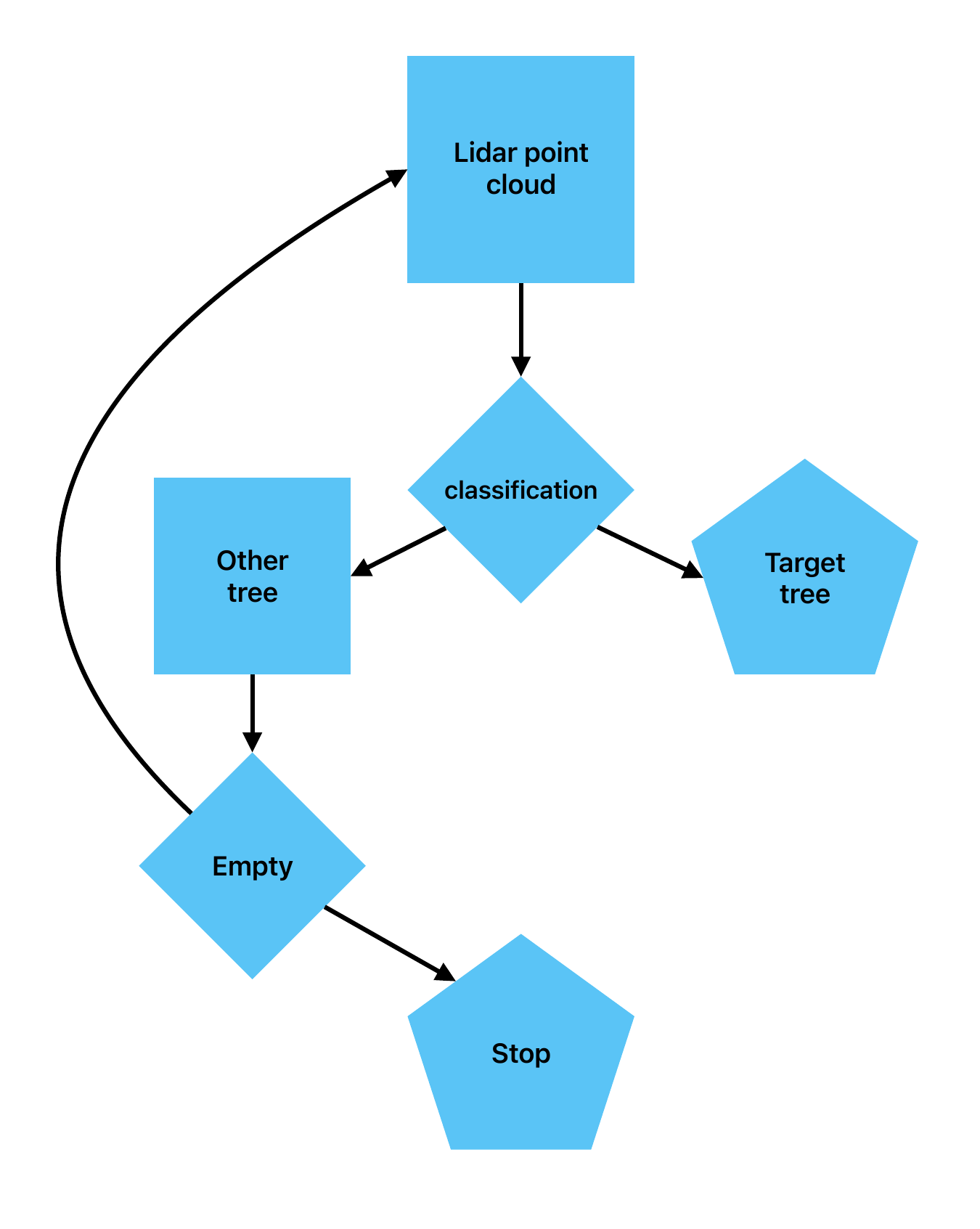}
    \caption{Pipeline of the region growing method.}

    \label{fig:pipeline_region_growing}
\end{figure}

The algorithm operates directly on a subset of the LiDAR point cloud classified as high vegetation.
Its core feature is an iterative point-wise assignment strategy: points are first sorted in descending order of elevation, and each pass of the algorithm extracts one tree, starting from the highest available point and proceeding downward.
Within a pass, every point is either assigned to the tree currently being grown or left aside, according to proximity and structural continuity criteria; the points left aside form the input of the following pass.
The procedure can be summarised as follows:

\begin{enumerate}
    \item The highest point of the point cloud is taken as the top of a new tree.
    \item All the sorted points are iteratively examined to determine whether they belong to this tree:
    \begin{itemize}
        \item if the point belongs to this tree, it is added to the tree ensemble;
        \item otherwise, it is temporarily left unassigned;
    \end{itemize}
    \item At the end of the pass, every point either belongs to the tree or is unassigned.
    \item The next pass takes only the unassigned points as its input point cloud and segments a new tree.
\end{enumerate}
This continues until every point has been assigned to a tree.
The corresponding pipeline is shown in Fig.~\ref{fig:pipeline_region_growing}.

To determine whether a point belongs to the current tree or not, we have followed these criteria.
\begin{itemize}
    \item If the point is not a local maximum, it must lie below a previously analyzed point.
    In this case, it inherits the tree label of its nearest analyzed neighbor. 
    \item If the point is a local maximum, it may represent either the top of a new tree or a branch of an existing one.
    We evaluate the local neighborhood: 
    \begin{itemize}
        \item if the point lies within a sufficiently close range to the current tree's top and its surrounding distribution is elongated (suggesting a branch-like structure), the point is assigned to the current tree. 
        \item otherwise, it may initiate a new tree in subsequent iterations. 
    \end{itemize}
\end{itemize}

\begin{figure}[H]
    \centering
    \includegraphics[width=0.49\textwidth]{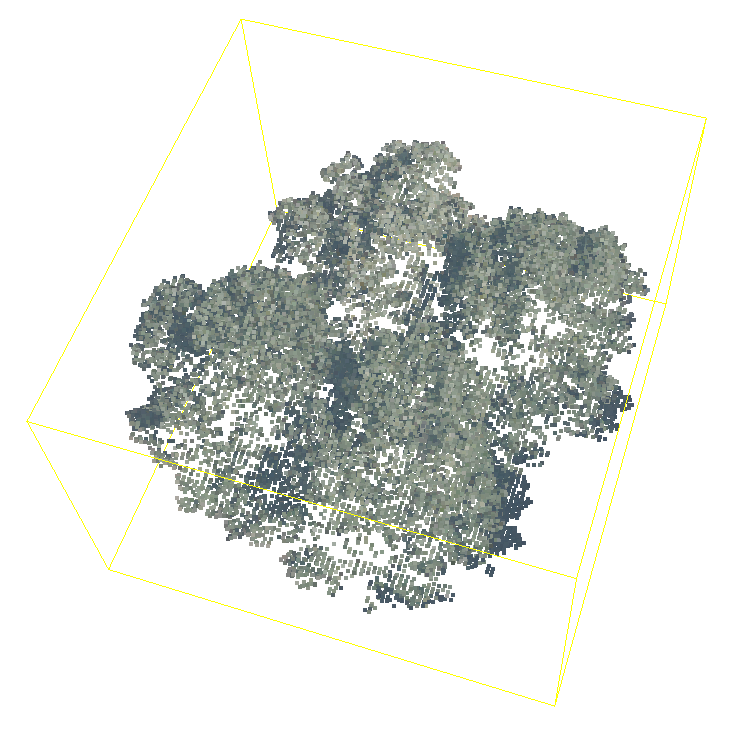}
    \includegraphics[width=0.49\textwidth]{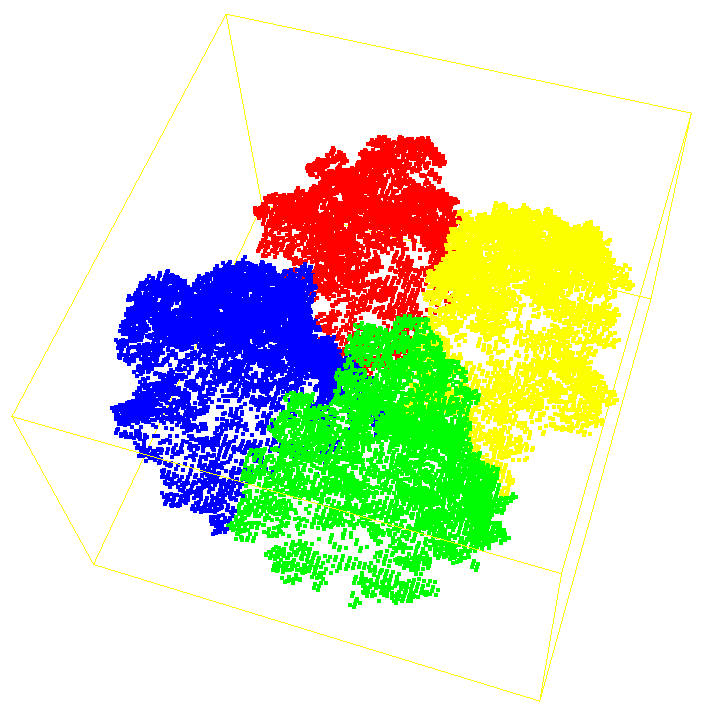}
    \caption{Example of segmentation using the region-growing algorithm.
    Four trees are correctly detected in the scene.}
    \label{fig:example_region_growing_1}
\end{figure}

\begin{figure}[H]
    \centering
    \includegraphics[width=0.49\textwidth]{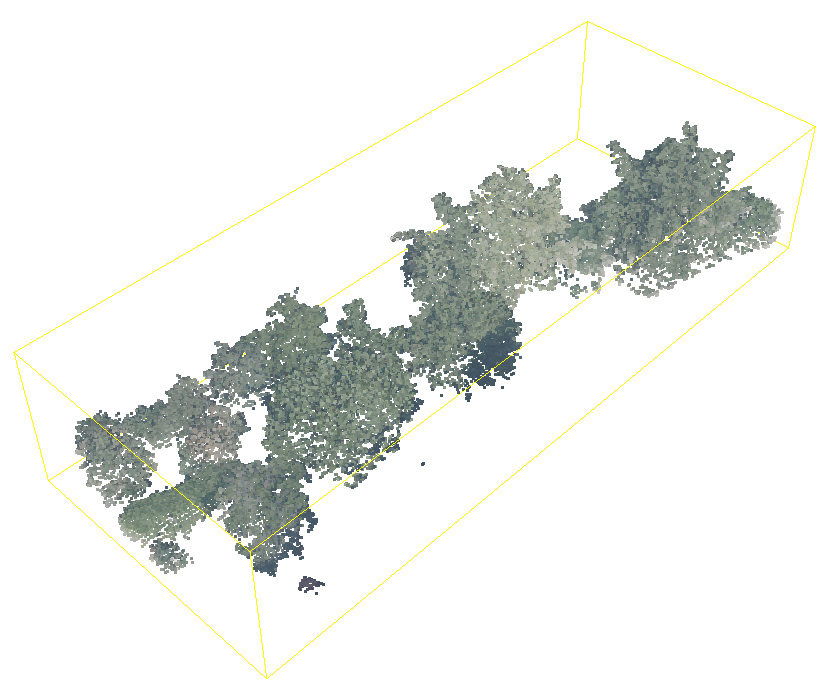}
    \includegraphics[width=0.49\textwidth]{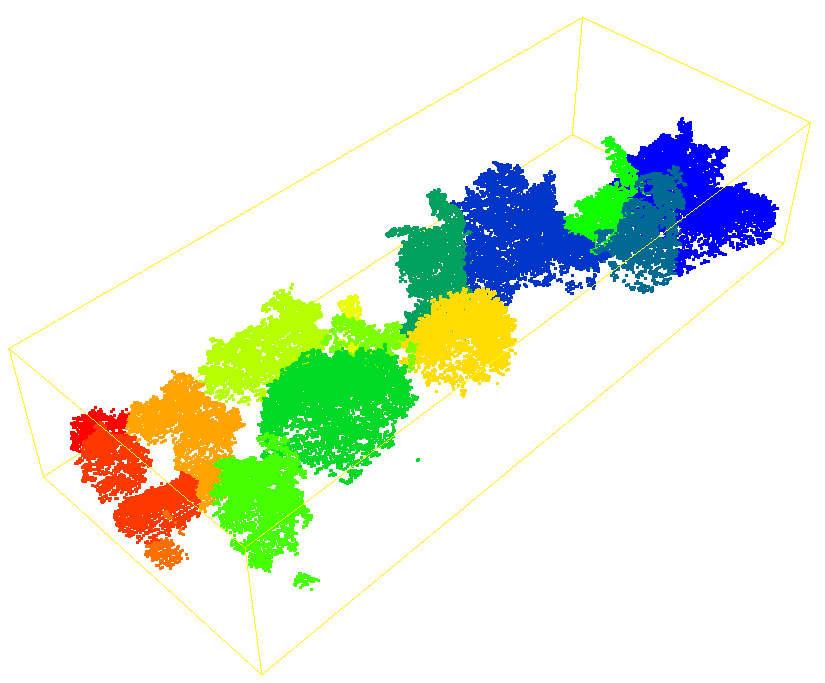}
    \caption{Example of segmentation using the region-growing algorithm on a more complex tree cluster.}
    \label{fig:example_region_growing_2}
\end{figure}

Some examples are depicted in Figs.~\ref{fig:example_region_growing_1} and~\ref{fig:example_region_growing_2}.
We can see how the algorithm identifies trees correctly, even though it slightly oversegments trees that possess elongated structures when faced with the problem of separating more complex clusters.
We are currently working on enforcing other controls over branch-like structures, in order to improve segmentation accuracy even further.  

\subsection{Numbers of trees, height and crown area comparison}
As part of the analysis, we generated four histograms -- the distributions of tree height and of crown radius, for each of the two segmentation methods -- using data collected from 12
spatial tiles of the 2023 campaign:
\begin{itemize}
    \item $32\_685000\_4929500$;
    \item $32\_685000\_4930000$;
    \item $32\_685000\_4930500$;
    \item $32\_685500\_4929500$;
    \item $32\_685500\_4930000$;
    \item $32\_685500\_4930500$;
    \item $32\_686000\_4929500$;
    \item $32\_686000\_4930000$;
    \item $32\_686000\_4930500$;
    \item $32\_686500\_4929500$;
    \item $32\_686500\_4930000$;
    \item $32\_686500\_4930500$.
\end{itemize}

\begin{figure}[H]
    \centering
    \includegraphics[width=0.8\textwidth]{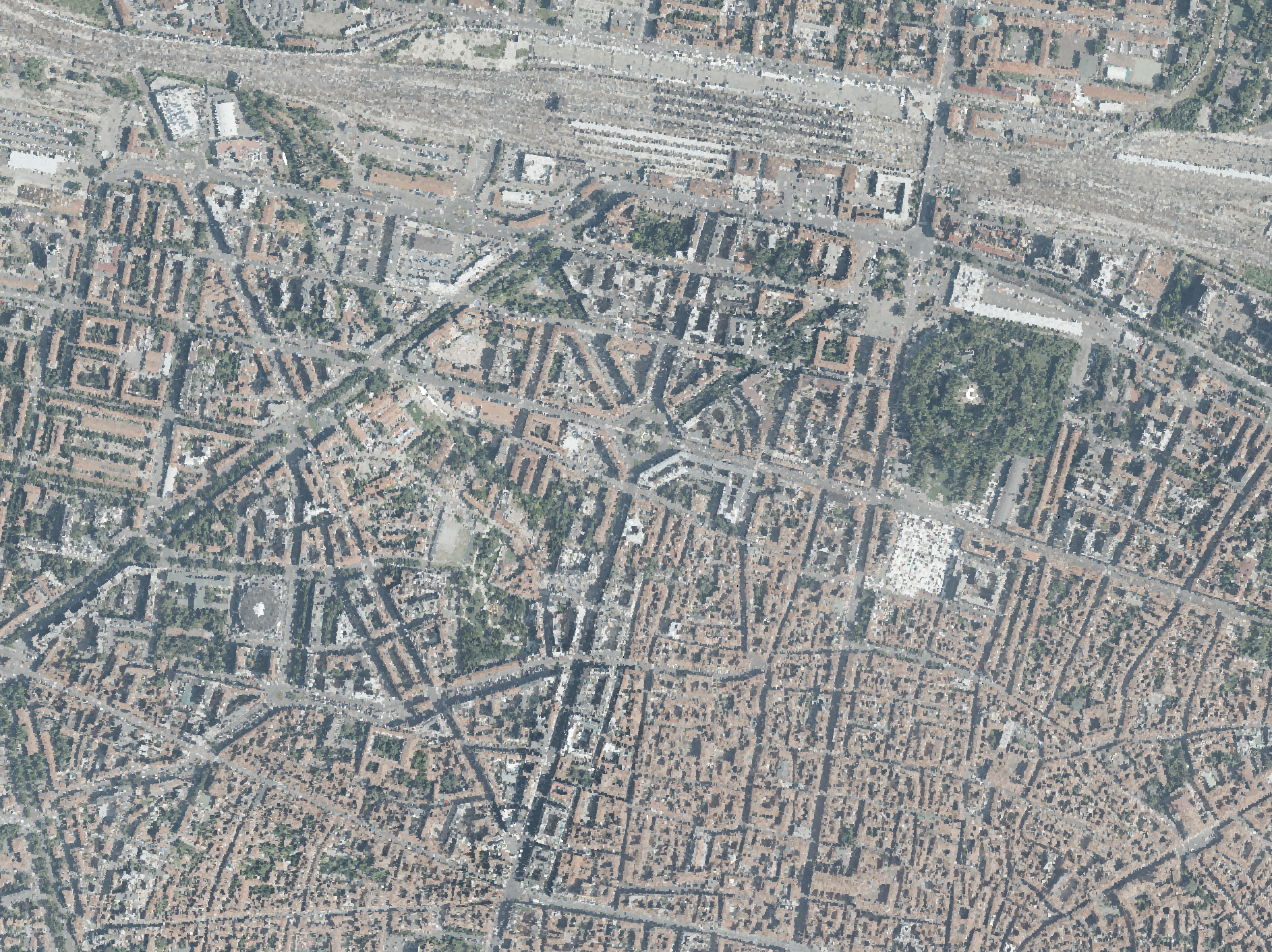}
    \caption{Extension of the area of Talea formed by merging the 12 tiles above.
    It is the area of \emph{Montagnola}, \emph{via Indipendenza}, \emph{Stazione Centrale} and \emph{viale Pietramellara} and \emph{Silvani}.}
    
    \label{fig:merge_tiles_talea}
\end{figure}

\begin{figure}[H]
    \centering
    \includegraphics[width=0.49\textwidth]{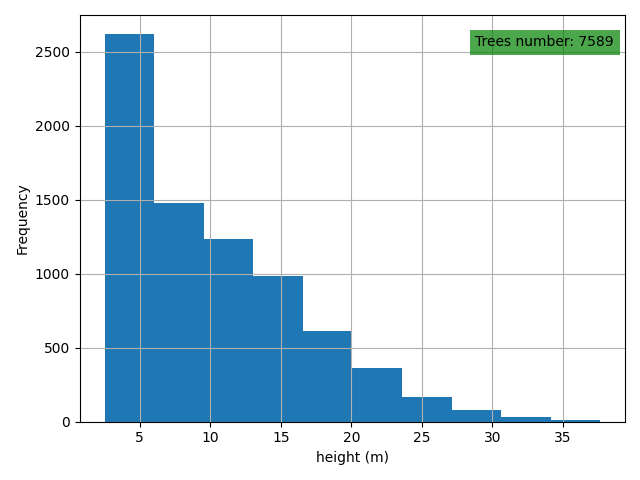}
    \includegraphics[width=0.49\textwidth]{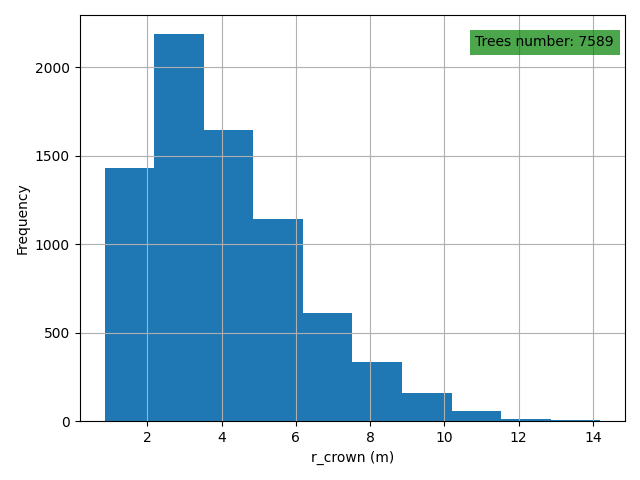}
    \caption{Left: Histogram trees height; Right: Histogram trees crown radius.
    Both histograms are based on data from 12 spatial tiles and derived from watershed analysis.}
    
    \label{fig:histograms_watershed}
\end{figure}

\begin{figure}[H]
    \centering
    \includegraphics[width=0.49\textwidth]{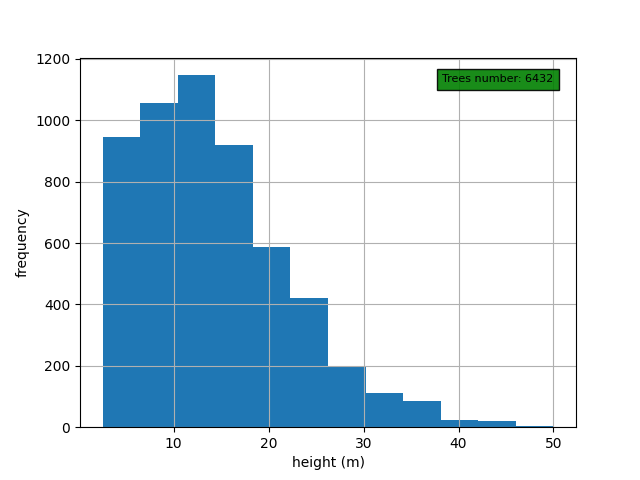}
    \includegraphics[width=0.49\textwidth]{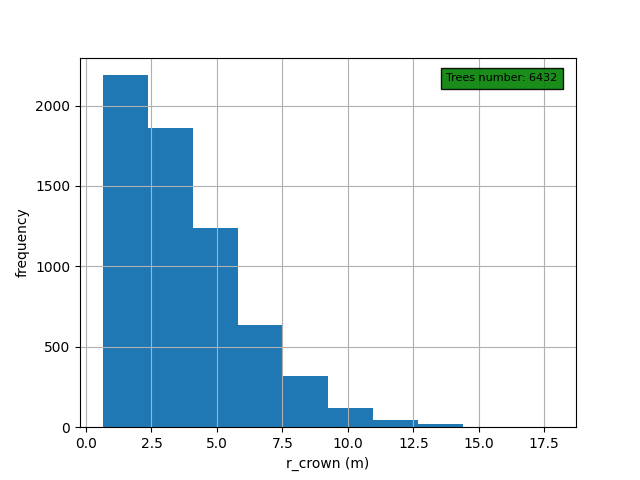}
    \caption{Left: Histogram trees height; Right: Histogram trees crown radius.
    Both histograms are based on data from 12 spatial tiles and derived from region growing analysis.}

    \label{fig:histogram_region_growing}
\end{figure}

The extent of the resulting area is shown in Fig.~\ref{fig:merge_tiles_talea}.
In each figure the left panel represents the distribution of tree heights $H$ and the right panel the distribution of crown radii $R$, computed through watershed (Fig.~\ref{fig:histograms_watershed}) and region growing (Fig.~\ref{fig:histogram_region_growing}).

Each tile was processed to extract individual tree metrics, and the aggregated values
across all $12$ tiles were used to compute the overall distributions. These histograms
provide a visual summary of the structural characteristics of the tree population over
the surveyed area.

The height histogram helps to identify the dominant vertical structure of the
vegetation, while the crown radius histogram gives insight into horizontal
spread, which is crucial for biomass estimation and ecological modeling.
The crown radius $R$ is defined as the radius of the circle having the same area as the tree canopy, so that it does not depend on the shape of the foliage.

\subsection{Comparison of the two methods}
\label{sec:comparison}
The comparison between the watershed and region growing segmentation methods reveals marked differences in the resulting distributions of both tree height and crown radius (see Figs.~\ref{fig:histograms_watershed} and~\ref{fig:histogram_region_growing}).
The watershed approach yields a higher total number of detected trees ($7589$), but the majority of these are relatively short, with a peak around 5 meters in height.
This overrepresentation of small trees is likely a consequence of the intrinsic characteristics of the watershed algorithm, which operates on a smoothed Canopy Height Model (CHM).
The smoothing process can lead to the fragmentation of continuous canopies into multiple local maxima, especially in densely vegetated or structurally complex areas, resulting in over-segmentation and the detection of several small objects rather than a smaller number of taller trees.

In contrast, the region growing method identifies a smaller total number of trees ($6432$), but it detects significantly taller individuals, with heights extending beyond 40 meters.
This suggests that region growing produces fewer small and short segments than watershed, since the number of trees taller than 10 meters is almost the same for the two methods.
In the absence of a smoothing step, region growing preserves the true elevation of the topmost returns, which is why individuals taller than 40 meters are detected.
By relying on proximity-based similarity and connectivity criteria rather than peak detection, it tends to preserve the full structure of taller trees and avoids their artificial fragmentation.

The analysis of crown radius distributions further supports this interpretation:
the two methods peak at almost the same value, between 2 and 3 meters of radius, but region growing detects wider crowns than watershed.

Additional factors contributing to these differences include the sensitivity of the watershed algorithm to CHM resolution and noise, the impact of local maxima detection strategies, and variations in post-processing steps such as minimum area filtering and merging of adjacent segments.

It is important to emphasize, however, that a rigorous quantitative evaluation of these segmentation methods is currently not feasible due to the lack of reliable ground truth data at the single-tree level.
As discussed earlier, the available Open Data catalogue is outdated and incomplete.
Addressing this limitation will require the development of curated reference datasets—either through collaborative field campaigns for direct data collection, or through the integration of complementary sensing technologies such as terrestrial (mobile) LiDAR and hyperspectral LiDAR, which can provide detailed structural and spectral information to support ground truth annotation.

\section{Vegetation Indicators}
Vegetation indicators are numerical metrics designed to estimate key ecological, structural, or physiological properties of urban green infrastructure. These indices play a fundamental role in quantifying the contribution of vegetation to ecosystem services—such as air quality improvement, carbon sequestration, temperature regulation, and biodiversity support—within complex urban environments.

In this article, we present a preliminary exploration of selected vegetation indicators derived from LiDAR-based tree segmentation and Open Data attributes. Our goal is not to propose a finalized system, but rather to introduce a set of computational tools and methodologies that may serve as a foundation for more refined and operational indicators in the future.

This initial phase should be understood as part of a broader process aimed at supporting municipal planning. As the Municipality of Bologna moves toward defining a comprehensive strategy for urban greenery—encompassing tree planting, maintenance, ecological goals, and citizen well-being—having a robust, data-driven framework for indicator computation will become essential. The tools introduced here are designed with adaptability in mind: they can be recalibrated, extended, or integrated with additional data sources (e.g., mobile LiDAR, hyperspectral imaging, crowdsourced observations) as the strategic vision evolves.

Two such indicators are presented in the following sections: potential carbon storage and allergenic potential. Both are built on the structural proxies extracted from the segmentation, namely the tree height and crown radius distributions discussed above. While approximate at this stage due to data limitations and the absence of full ground truth validation, these indicators demonstrate the feasibility and relevance of systematic, scalable analysis pipelines that can directly inform urban forestry policies.
\subsection{Carbon Storage Index}
\label{sec:carbon}

Urban trees play a fundamental role in mitigating climate change by capturing and storing atmospheric carbon dioxide (CO\textsubscript{2}) through the process of photosynthesis. This mechanism, known as \textit{carbon sequestration}, leads to the accumulation of carbon in the tree's above-ground biomass (AGB), which includes the trunk, branches, and foliage.

To estimate the amount of carbon stored in each individual tree, we adopt an allometric modeling approach that leverages structural parameters derived from remote sensing data. Specifically, we use tree height (\( H \)) and crown radius (\( R \)), both of which are extracted from the LiDAR-based individual tree segmentation results. These parameters serve as inputs to empirical allometric equations that relate tree geometry to biomass.

When species information is available—either from field surveys or ancillary datasets such as the municipal tree inventory—it is used to select species-specific or genus-specific allometric coefficients. This allows for a more accurate estimation of biomass and, consequently, of stored carbon. In the absence of detailed species data, generalized allometric models based on broadleaf or conifer classifications can be applied as a first approximation.

We rely on the model proposed by \cite{lin2023precise}, which expresses AGB as
a function of height and crown area:

\begin{equation}
\label{eq:agb}
\text{AGB} = \alpha \cdot \left[ H \cdot \ln(\pi R^2) \right]^\beta
\end{equation}

where AGB is expressed in kg, the tree height $H$ in meters and the crown area $\pi R^2$ in square meters.
For the coefficients we adopt the values fitted in~\cite{lin2023precise}, namely
\[
\alpha = 0.0511, \qquad \beta = 1.9486,
\]
obtained by destructive sampling against UAV-borne LiDAR measurements with a coefficient of determination $R^2 = 0.9477$ and a root mean square error of $6.04$~kg.

Two caveats must be stated explicitly.
First, these coefficients were calibrated on a plantation of dawn redwood (\textit{Metasequoia glyptostroboides}), a conifer with a regular conical crown, whereas the urban forest of Bologna is dominated by broadleaves with markedly different crown architectures; applying them city-wide is therefore a first-order approximation.
Second, Eq.~\eqref{eq:agb} is not defined for crowns smaller than $1/\sqrt{\pi} \approx 0.56$~m in radius, since the logarithm of the crown area becomes negative; such segments are excluded from the carbon computation.
A species-resolved calibration is the natural next step, and the Urban Tree Database of McPherson et al.~\cite{mcpherson2016urban, mcpherson2016urbandata} provides a suitable basis for it: it collects field measurements of more than $14\,000$ street and park trees across 171 species, together with the corresponding volume and dry-weight biomass equations, and is explicitly designed for urban rather than forest stands.
Recalibrating $\alpha$ and $\beta$ per genus against that dataset -- or against a local destructive-sampling campaign -- would remove the largest single source of systematic error in the present carbon estimates.

The above-ground biomass is then converted into carbon using a standard coefficient of 0.47, following IPCC (Intergovernmental Panel on Climate Change) guidelines:
\[
\text{C} = \text{AGB} \cdot 0.47
\]

Finally, carbon is converted to the equivalent amount of 
CO\textsubscript{2} using the molecular weight ratio:

\begin{align*}
    \text{CO}_2 &= C \cdot \frac{\text{atomic weight of 1 atom of C} + \text{atomic weight of 2 atoms of O}}{\text{atomic weight of 1 atom of C}} \\\\
    &= C \cdot \frac{12 + 2 \cdot 16}{12} = C \cdot \frac{44}{12} = C \cdot 3.67
\end{align*}

\[
\text{CO\textsubscript{2}} = \text{C} \cdot 3.67 = \text{AGB} \cdot 1.7249
\]

\subsubsection{Data Sources}
\label{sec:data_sources}

\begin{itemize}
    \item Watershed lidar tree polygons shapefile: A geospatial file containing the segmented crown geometries of individual trees.
    \item OpenData CSV file: Contains point data with latitude, longitude, and the tree species (column \texttt{Specie arborea}).
\end{itemize}

The indicators presented in this section are computed on the watershed segmentation rather than on the region growing one.
This choice is dictated by the nature of the two outputs.
The watershed algorithm returns closed two-dimensional crown polygons in map geometry, which can be intersected directly with the georeferenced points of the municipal catalogue to transfer the species label, and whose area provides the crown area $A = \pi R^2$ required by the allometric model without any further reconstruction step.
The region growing algorithm, on the other hand, returns a per-point tree label in three dimensions: obtaining an explicit crown boundary from it requires an additional projection and hull-fitting stage, which would introduce its own parameters and uncertainties into the indicator estimates.
Moreover, the over-segmentation discussed in Section~\ref{sec:comparison} affects almost exclusively the low end of the height distribution, and since above-ground biomass scales steeply with $H$, the spurious small segments contribute a negligible fraction of the total estimated carbon stock.
Their effect on the pollen estimate is potentially larger, because Eq.~\eqref{eq:crown_volume} is dominated by the crown radius, and this is one of the reasons why the allergen map should be read as a relative, rather than absolute, indicator.
Extending the indicator computation to the region growing output, once a robust crown reconstruction is in place, is left to future work.

\subsubsection{Method Overview}
This methodology allows us to estimate carbon storage at the individual tree 
level and, by aggregating results, to generate high-resolution spatial layers 
of urban carbon stock across the entire city.

This script performs a spatial join between a CSV file containing tree species 
and geographic coordinates, and a shapefile with pre-segmented tree crown polygons. Its goal is to assign species information to each polygon by matching it with the closest point in the CSV file, which represents individual trees.

\begin{enumerate}
    \item The shapefile is loaded and reprojected to WGS84 (EPSG:4326). Two new columns are initialized: one to store the assigned species, and one boolean flag (\texttt{assigned}) to avoid assigning multiple species to the same polygon.
    
    \item The CSV file is read and converted to a \texttt{GeoDataFrame} using the \texttt{Geo Point} column, which stores coordinates in the format \texttt{latitude,longitude}.
    
    \item A spatial index is built on the polygons to enable efficient nearest-neighbor queries.
    
    \item For each point (tree with known species), the algorithm:
    \begin{enumerate}
        \item Searches for polygons within a 30-meter buffer;
        \item If no candidates are found, it falls back to considering all polygons;
        \item Filters out polygons that have already been assigned a species;
        \item Computes the distance between the point and the centroid of each candidate polygon;
        \item Selects the closest polygon and assigns the species name to it.
    \end{enumerate}
    
    \item The updated \texttt{GeoDataFrame} is then saved as a new shapefile, with the assigned species stored in the \texttt{specie} column.
\end{enumerate}



\begin{figure}[H]
    \centering
    \includegraphics[width=0.8\textwidth]{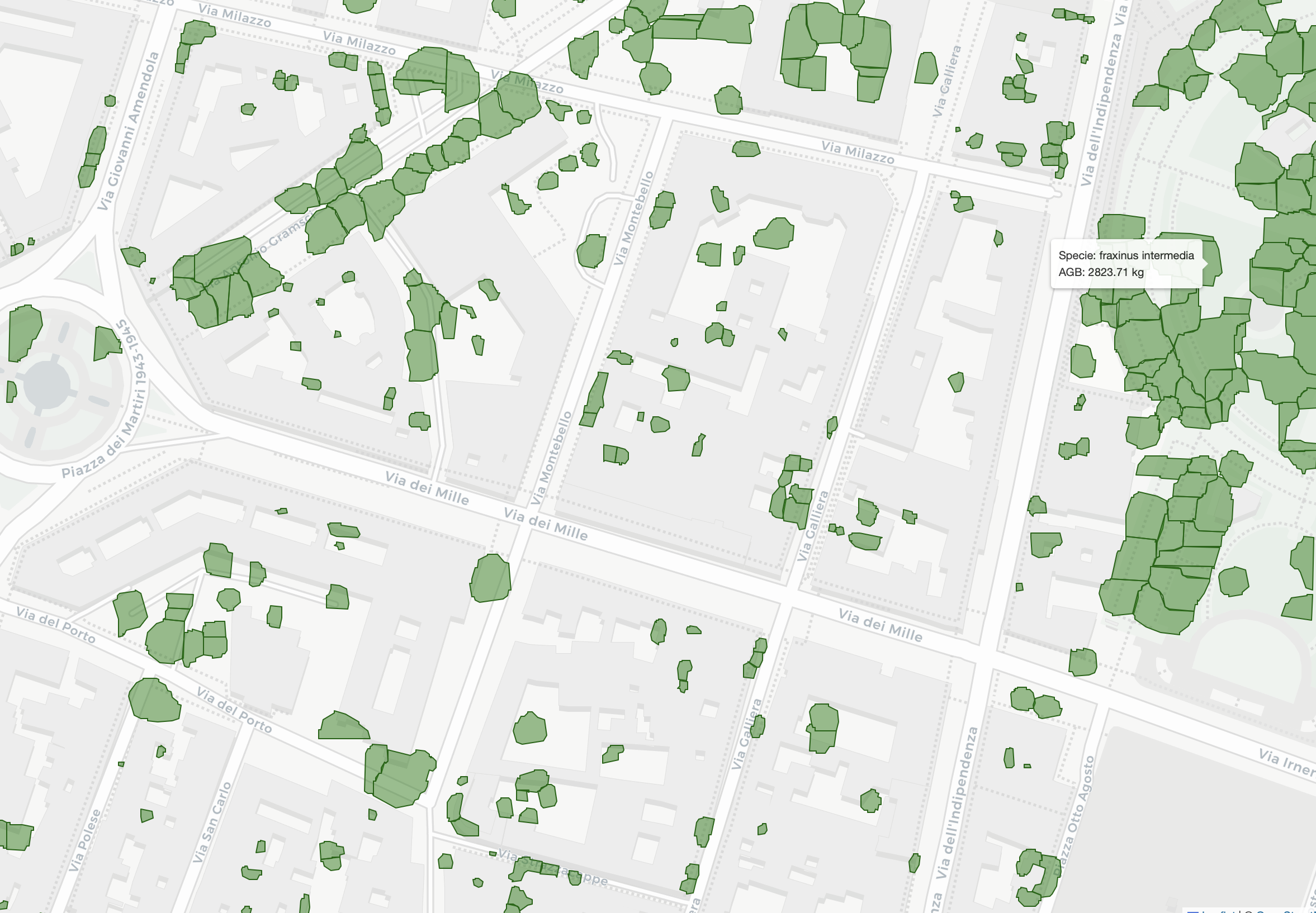}
    \caption{AGB estimation using Lin approach.}
    
    \label{fig:agb_estimation}
\end{figure}

The output is a shapefile where each segmented crown polygon has a species label assigned based on proximity to known tree locations. This allows for downstream processing such as biomass or pollen estimation using species-specific parameters.
The per-tree above-ground biomass obtained by applying Eq.~\eqref{eq:agb} to the segmented crowns is mapped in Fig.~\ref{fig:agb_estimation}.

\subsection{Allergen Index}
Airborne allergens are microscopic biological particles, such as pollen grains, that can 
trigger allergic reactions in sensitive individuals. Among the most common sources are 
tree species, whose pollen production varies depending on their morphology and 
reproductive strategy. To estimate the annual pollen output of urban trees, we rely on 
an empirical formula:

\begin{equation}
\label{eq:pollen_production}
P = D \cdot V_{\text{crown}} \cdot E
\end{equation}

where $P$ is the number of pollen grains produced per year, $D$ is the density of
inflorescences per cubic meter of crown volume, $V_{\text{crown}}$ is the estimated crown
volume, and $E$ is the average number of pollen grains per inflorescence. The crown volume
is approximated as a spheroid built on the tree height and crown radius, as detailed below.

This method is supported by peer-reviewed studies such as \cite{Katz2020pollen}, which 
provides species-specific coefficients for common urban trees. Additional reference 
values were adapted from \cite{peper2001predictive} and \cite{peper2001equations}. 
The coefficients $D$ and $E$ used in this analysis were compiled into a reference table 
based on these sources and include representative species such as \textit{Betula pendula}, 
\textit{Quercus robur}, and \textit{Platanus acerifolia}.

The tree species recorded in the shapefile are compared against those listed in the
coefficient file using partial word matching; the input data are the same as in Section~\ref{sec:data_sources}.
For instance, species names like \textit{Platanus acerifolia} are matched to entries in
the coefficient dataset even if only part of the name (e.g., ``Platanus'') appears. This
approach improves robustness against naming variations or partial inputs.


Each tree crown is approximated as a spheroid whose two horizontal semi-axes are equal to the crown radius $R$ and whose vertical semi-axis is half the crown depth $L$, that is, half the vertical extent of the foliage between the crown base and the treetop.
Ideally $L$ would be measured directly on the segmented point cloud, as the difference between the elevation of the treetop and that of the lowest returns belonging to the crown.
This is not feasible with the present data: the acquisition is airborne, so the sensor illuminates the canopy from above and the returns are concentrated on the upper, outer envelope of the crown.
The lower branches and the crown base are shadowed by the foliage above them, and in the denser urban canopies of Bologna they receive few or no returns at all; what little is recorded below the crown is moreover hard to separate from understory shrubs, parked vehicles, street furniture and ground points.
A crown base extracted from such data would be driven by the local point density rather than by the actual tree architecture, and would vary from tile to tile with the flight geometry.
We therefore fix the crown depth to a constant fraction of the tree height, $L = H/2$, i.e. we assume a live crown ratio $L/H = 0.5$.
This choice reflects the composition of the urban forest analysed here: street and park trees are routinely pruned to raise the crown above the roadway and the sidewalk, so that a sizeable lower fraction of the trunk carries no foliage, and taking the crown to occupy the upper half of the tree is a deliberately conservative summary of that geometry.
The crown-height measurements collected in the Urban Tree Database~\cite{mcpherson2016urban, mcpherson2016urbandata} would allow this ratio to be calibrated per genus, which we leave to future work together with the recalibration of the allometric coefficients discussed in Sec.~\ref{sec:carbon}.
The alternative of letting the spheroid span the full height, $L = H$, would instead place foliage along the whole trunk and yield a systematic overestimate of the crown volume by a factor of two.
With $L = H/2$ the vertical semi-axis is $H/4$ and the general ellipsoid volume $V = \frac{4}{3}\pi\,abc$ gives

\begin{equation}
\label{eq:crown_volume}
V_{\text{crown}} = \frac{4}{3} \pi \, R \cdot R \cdot \frac{H}{4} = \frac{1}{3} \pi R^2 H,
\end{equation}

where $R$ is the average crown radius and $H$ is the tree height, both derived from the LiDAR segmentation as described above.
It should be stressed that Eq.~\eqref{eq:crown_volume} remains a first-order approximation: the crown ratio is treated as a single constant for all species and all individuals, whereas it varies with species, age and pruning regime, and the spheroid ignores the internal gaps of the foliage.
Replacing the constant with a species-specific or, better, an individually measured crown depth --- which would require a terrestrial or mobile LiDAR acquisition able to see the lower crown --- is a natural refinement of the present pipeline.
Equation~\eqref{eq:crown_volume} is then substituted into Eq.~\eqref{eq:pollen_production} to obtain the estimated number of pollen grains produced per tree per year.


Trees are excluded from the pollen estimation in the following cases:
\begin{enumerate}
    \item Trees with no recorded species name,
    \item Trees whose species name does not match any entry in the coefficient file (leading to missing $D$ and $E$),
    \item Trees without valid crown radius information.
\end{enumerate}

These trees are not visualized in the final pollen production map, as the required data for the computation is incomplete.

\begin{figure}[H]
    \centering
    \includegraphics[width=0.8\textwidth]{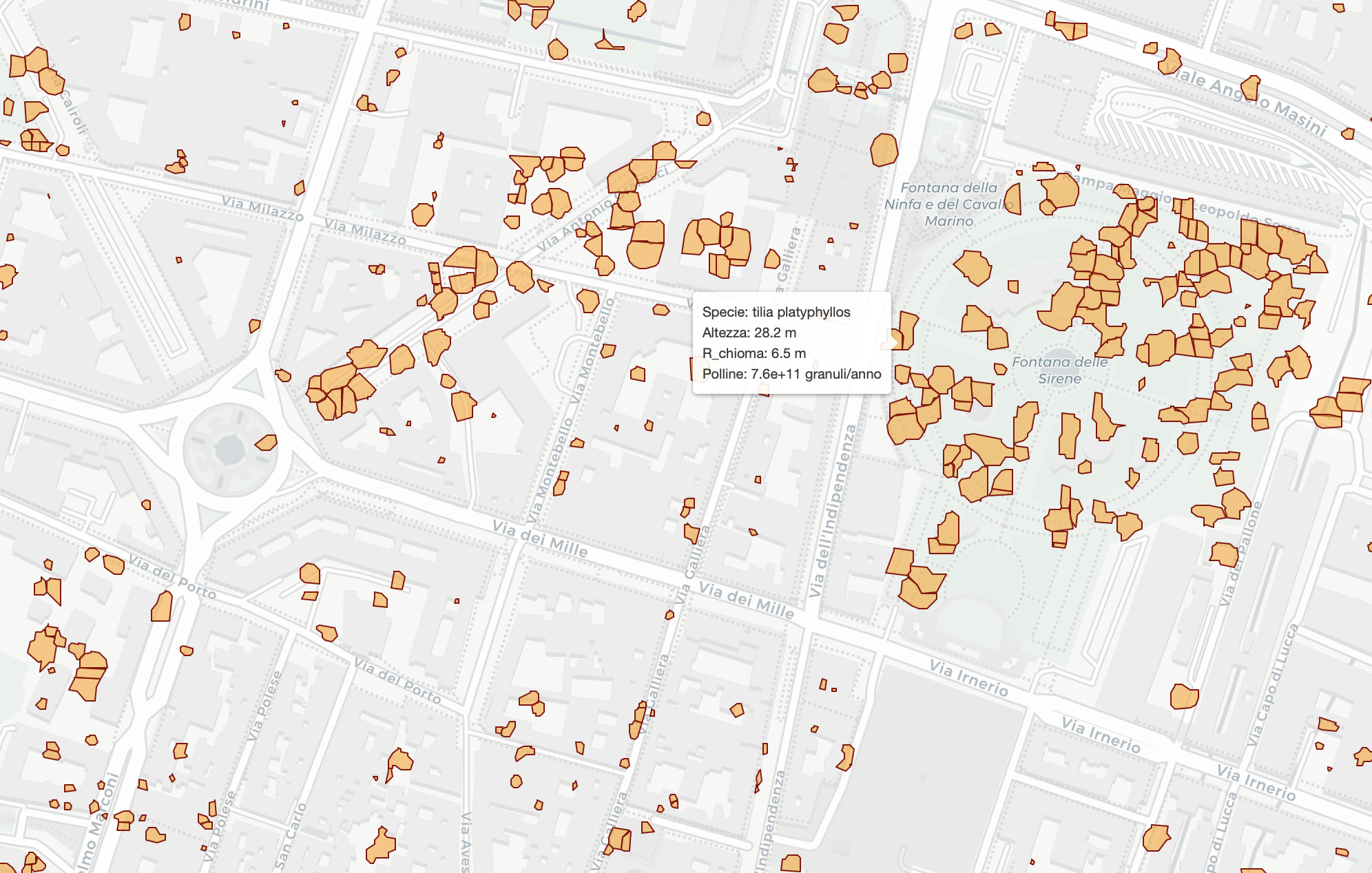}
    \caption{Tree allergens estimation using equation~\ref{eq:pollen_production}.}
    
    \label{fig:allergens_estimation}
\end{figure}

This approach enables spatially explicit estimates of allergenic pollen production in
urban areas, which is essential for assessing environmental exposure and guiding tree
management policies. The resulting map is shown in Fig.~\ref{fig:allergens_estimation}.

\section{Trees Metadata Map}
\begin{figure}[H]
    \centering
    \includegraphics[width=1\textwidth]{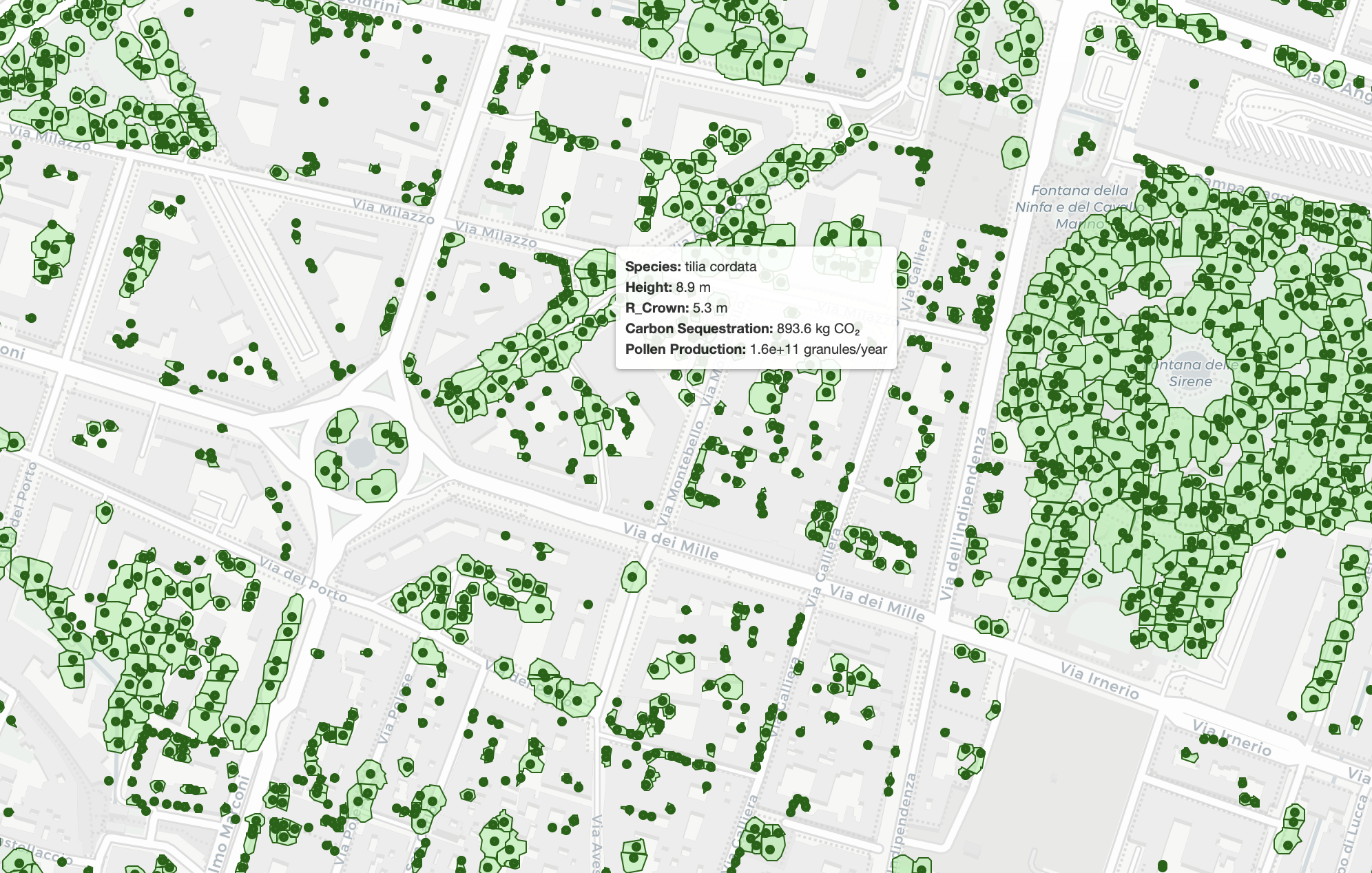}
    \caption{Trees metadata map.}
    
    \label{fig:metadat_map}
\end{figure}

All the information derived from this analysis has been integrated into an interactive map
built with \texttt{folium}, distributed together with the code as
\href{https://github.com/physycom/Vegetation_Segmentation/blob/main/report/link/mappa_carbonio_polline.html}{\textcolor{blue}{\texttt{report/link/mappa\_carbonio\_polline.html}}},
to be downloaded and opened in a browser.
The map displays metadata for each individual tree, including its
shape (polygon footprint), treetop location (point), height, crown radius 
(stored in the \texttt{r\_crown} column), species, estimated carbon sequestration, and annual pollen production.
This spatial visualization, illustrated in Fig.~\ref{fig:metadat_map}, enables a comprehensive inspection of urban trees, combining
morphological and ecological indicators.

\section{Conclusions}
In this work we assembled and tested an end-to-end pipeline that goes from raw airborne LiDAR acquisitions to per-tree ecological indicators for the urban vegetation of Bologna.

Two individual-tree segmentation strategies were implemented and compared.
The watershed algorithm, operating on a smoothed Canopy Height Model, is computationally light and provides well-defined crown polygons that can be directly used to filter the point cloud, but it tends to over-segment dense or structurally complex canopies, producing a large population of small, low objects.
The region growing algorithm works directly on the three-dimensional points and better preserves the vertical continuity of tall trees, at the cost of a slight over-segmentation of individuals with elongated branch structures.
The two methods agree on the number of trees above 10 meters and on the position of the crown radius peak, so their disagreement is concentrated in the low and small end of the distributions, exactly where the smoothing step of the watershed approach is most influential.

The comparison with the municipal Open Data catalogue showed that the currently available inventory cannot serve as a ground truth: on the tile examined in detail about half of the LiDAR-detected trees are missing from it, several catalogued geopoints correspond to locations where no tree is present, and the majority of the height records are roughly twenty years old.
Extending this comparison to the full set of tiles is a straightforward next step and would turn a single-tile observation into a city-scale figure.
This is the main limitation of the present study, since it prevents a rigorous quantitative validation of either segmentation method.
It also indicates, on the other hand, one of the most immediate practical benefits of the proposed pipeline: periodic aerial LiDAR acquisitions can be used to keep the municipal inventory aligned with the actual state of the urban forest, updating positions, heights and crown extents and flagging trees that have been removed or newly planted.

The structural features extracted from the segmented trees were used to compute two preliminary indicators, carbon storage through allometric biomass models and annual pollen production through species-specific inflorescence coefficients, and the results were integrated into an interactive metadata map.
These indicators should be regarded as demonstrators rather than as final products: their absolute values depend on allometric and species coefficients calibrated elsewhere, on species labels inherited from the Open Data catalogue through a nearest-neighbour spatial join, and on a crown depth fixed to half the tree height for all individuals.

Future developments will follow three directions.
First, improving the segmentation itself, in particular by enforcing additional controls on branch-like structures within the region growing algorithm.
Second, building a curated reference dataset, either through dedicated field campaigns or through complementary acquisitions such as terrestrial or mobile LiDAR and hyperspectral sensors, which would also enable species identification and health assessment beyond what aerial platforms can provide.
Third, refining and extending the set of indicators in agreement with the strategic priorities defined by the Municipality.
The modular design of the pipeline means that each of these components can be replaced or recalibrated independently, so that the framework can evolve together with the data and with the policy questions it is meant to support.

\newpage
\printbibliography[nottype=misc, title={Bibliography}]
\printbibliography[type=misc, title={Websites Resources}]

\end{document}